\documentclass[sort&compress,times,fleqn]{elsarticle} 
\usepackage{graphicx}
\usepackage{color}
\usepackage{enumerate}
\usepackage[breaklinks]{hyperref}
\hypersetup{colorlinks,urlcolor=blue,citecolor=red,linkcolor=blue}
\usepackage[latin1,utf8]{inputenc}
\usepackage[OT2,OT1]{fontenc}

\usepackage{latexsym}
\usepackage{amssymb,amsthm}
\usepackage{bm}

\usepackage{tikz}
\usepackage[compat=1.1.0]{tikz-feynman}
\usepackage{silence} 
\usepackage{orcidlink} 
\usepackage[makeroom]{cancel}
\usepackage{enumitem}
\usepackage{amssymb,amsmath,mathrsfs,bm,bbm,mathtools} 
\usepackage{amsthm}
\definecolor{ao(english)}{rgb}{0.0, 0.5, 0.0}
\hypersetup{colorlinks=true,linkcolor=blue,urlcolor=blue,citecolor=ao(english)}
\usepackage{stackrel}
\usepackage{multirow}
\usepackage[normalem]{ulem} 

\newcommand\nn{\nonumber\\}

\newcommand{\bma}{\left(\begin{array}}
	\newcommand{\ema}{\end{array}\right)}
\newcommand{\be}{\begin{equation}}
	\newcommand{\ee}{\end{equation}}
\newcommand{\ben}{\begin{equation*}}
	\newcommand{\een}{\end{equation*}}
\newcommand{\ba}{\begin{eqnarray}}
	\newcommand{\ea}{\end{eqnarray}}
\newcommand{\ban}{\begin{eqnarray*}}
	\newcommand{\ean}{\end{eqnarray*}}
\newcommand{\bs}{\begin{subequations}}
	\newcommand{\es}{\end{subequations}}
\newcommand{\bc}{\begin{center}}
	\newcommand{\ec}{\end{center}}
\newcommand{\ve}{\varepsilon}

\def\arg{{\rm arg}}
\def\Arg{{\rm Arg}}

\newcommand{\Bd}{\boxdot}

\newcommand{\au}[2]{#1.~#2}
\newcommand{\arX}[1]{\href{http://arxiv.org/abs/#1}{{\cob arXiv:#1}}}
\newcommand{\oarX}[1]{\href{http://arxiv.org/abs/#1}{{\cob #1}}}

\newcommand{\book}[5]{\emph{#1}, #2, #3, #4 (#5)}
\newcommand{\books}[4]{\emph{#1}, #2, #3 (#4)} 
\newcommand{\doin}[6]{\href{http://dx.doi.org/#1}{{\cob {\it #2 #3} {\bf #4} (#6) #5}}}
\newcommand{\doinn}[5]{\href{http://dx.doi.org/#1}{{\cob {\it #2} {\bf #3} (#5) #4}}}
\newcommand{\doij}[5]{\href{http://dx.doi.org/#1}{{\cob {\it #2} {\bf #3} (#5) #4}}}
\newcommand{\ndoin}[6]{\href{#1}{{\cob {\it #2 #3} {\bf #4} (#6) #5}}}
\newcommand{\ndoinn}[5]{\href{#1}{{\cob {\it #2} {\bf #3} (#5) #4}}}

\newcommand{\tia}[1]{\textit{#1},}

\newcommand{\boxd}[1]{\boxed{\phantom{\Biggl(}#1\phantom{\Biggl)}}}

\renewcommand{\leq}{\leqslant}
\renewcommand{\geq}{\geqslant}
\newcommand{\Eq}[1]{(\ref{#1})}
\newcommand{\Eqq}[1]{Eq.~(\ref{#1})}
\newcommand{\Eqqs}[1]{Eqs.~(\ref{#1})}
\def\rme{e}
\def\rmd{d}
\def\rmi{i}

\def\Re{\text{Re}}
\def\Im{\text{Im}}
\def\erf{{\rm erf}}
\def\a{\alpha}
\def\b{\beta}
\def\de{\delta}

\def\g{\gamma}
\def\la{\lambda}

\def\e{\epsilon}
\def\ve{\varepsilon}

\def\om{\omega}
\def\G{\Gamma}
\def\t{\tau}
\def\s{\sigma}
\def\vr{\varrho}
\def\vp{\varphi}
\def\N{\nabla}
\def\B{\Box}

\def\lst{\ell_*}

\def\cC{\mathcal{C}}
\def\cD{\mathcal{D}}

\def\cF{\mathcal{F}}
\def\cG{\mathcal{G}}
\def\cH{\mathcal{H}}

\def\cK{\mathcal{K}}
\def\cL{\mathcal{L}}

\def\cS{\mathcal{S}}

\def\p{\partial}

\def\cob{\color{blue}}

\newtheorem*{theo}{Theorem}

\begin{document}
	
\begin{frontmatter}

\title{Representations of the fractional d'Alembertian and initial conditions in fractional dynamics} 

\author{Gianluca Calcagni\,\orcidlink{0000-0003-2631-4588}\corref{cor1}}
\ead{g.calcagni@csic.es}
\address{Instituto de Estructura de la Materia, CSIC, Serrano 121, 28006 Madrid, Spain}
\cortext[cor1]{Corresponding author.}
		
\author{Giuseppe Nardelli\,\orcidlink{0000-0002-7416-6332}}
\ead{giuseppe.nardelli@unicatt.it}
\address{Dipartimento di Matematica e Fisica, Universit\`a Cattolica del Sacro Cuore, Via della Garzetta 48, 25133 Brescia, Italy}
\address{TIFPA -- INFN, c/o Dipartimento di Fisica, Universit\`a di Trento, 38123 Povo (TN), Italy}

\begin{abstract}
We construct representations of complex powers of the d'Alembertian operator $\Box$ in Lorentzian signature and pinpoint one which is self-adjoint and suitable for classical and quantum fractional field theory. This self-adjoint fractional d'Alembertian is associated with complex-conjugate poles, which are removed from the physical spectrum via the Anselmi--Piva prescription. As an example of empty spectrum, we consider a purely fractional propagator and its K\"all\'en--Lehmann representation. Using a cleaned-up version of the diffusion method, we formulate and solve the problem of initial conditions of the classical dynamics with a standard plus a fractional d'Alembertian, showing that the number of initial conditions is two. We generalize this result to a much wider class of nonlocal theories and discuss its applications to quantum gravity.	
\end{abstract}


\end{frontmatter}
	
\date{May 27, 2025}


\tableofcontents


\section{Introduction}\label{sec1}

Since the failure of standard quantum field theory (QFT) to renormalize Einstein gravity \cite{tHooft:1974toh,Deser:1974cz,Deser:1974cy,Deser:1974xq,Goroff:1985sz,Goroff:1985th,vandeVen:1991gw,Bern:2015xsa,Bern:2017puu}, much effort has been dedicated to find alternative ways to quantize this theory or one of its extensions \cite{Ori09,Fousp,Calcagni:2013hv,Mielczarek:2017cdp,Bam24}. However, in recent years a sizable portion of the community's attention has been redirected back to perturbative QFT and to ways of adapting its ingredients to gravity without spoiling its main tenets. For example, in nonlocal quantum gravity (NLQG) \cite{Modesto:2017sdr,Buoninfante:2022ild,BasiBeneito:2022wux,Koshelev:2023elc} one modifies the action with nonlocal operators in such a way as to achieve renormalizability and unitarity but at the price of replacing the usual analytic continuation of the amplitudes (Wick rotation) with Efimov analytic continuation \cite{Pius:2016jsl,Briscese:2018oyx,Chin:2018puw}. In another example, which we may call quantum gravity with fakeons \cite{Anselmi:2017yux,Anselmi:2017lia,Anselmi:2018bra,Anselmi:2019rxg,Anselmi:2021hab,Anselmi:2022toe,Anselmi:2025uzj}, one obtains renormalizability with higher-order curvature terms as in Stelle gravity \cite{Stelle:1976gc,Stelle:1977ry,Julve:1978xn,Fradkin:1981hx,Fradkin:1981iu,Avramidi:1985ki,Hindawi:1995uk,Hindawi:1995an} and unitarity is preserved by a Lorentz-invariant non-analytic prescription of Lorentzian amplitudes \cite{Anselmi:2025uzj}. In a third case, the geometry of spacetime and correlation functions change at different scales \cite{Calcagni:2016azd,Carlip:2019onx,Calcagni:2021ipd}. This promotion of spacetime to a deterministic or to a random multi-fractal can be done in different ways but, eventually, one was singled out as the most manageable because it preserves Lorentz invariance: fractional quantum gravity (FQG) \cite{Calcagni:2021aap,Calcagni:2022shb}. This is a nonlocal QFT just like NLQG but with different kinetic terms, based on non-integer (or, as commonly denoted, ``fractional'') powers of the d'Alembertian operator $\B\equiv \N_\mu\N^\mu$, where $\N_\mu$ is the covariant derivative in $D$ spacetime dimensions ($\mu=0,1,\dots,D-1$). The main message gathered from all these scenarios is that perturbative QFT \emph{is} a viable framework to quantize gravity, provided one is willing to tamper with technicalities to guarantee mathematical and physical consistency.

The example of FQG shows that fractional d'Alembertians can play an important role in QFT similar to higher-derivative local operators $\B^n$, as a way to improve renormalizability and eventually to obtain an ultraviolet-complete description of gravity. At the same time, the intrinsically different nature of the spacetime geometry realized with fractional d'Alembertians \cite{Calcagni:2016azd} may also be the one responsible for the different and perhaps more promising treatment of unitarity of these models with respect to higher-derivative ones. However, although FQG and fractional QFTs in general are power-counting renormalizable and there is evidence that they can be made unitary \cite{Calcagni:2022shb,Trinchero:2012zn,Trinchero:2017weu,Trinchero:2018gwe,Calcagni:2021ljs}, little is still known about their classical solutions. Since FQG is a nonlocal theory and contains an infinite number of derivatives, its Cauchy problem cannot be formulated in the usual way by taking the kinetic operator ``as is'' and counting a finite number of independent initial conditions on the field and its derivatives. This long-standing issue in nonlocal dynamics \cite{Pais:1950za,Eliezer:1989cr,Moeller:2002vx} has been solved for NLQG by using the diffusion method \cite{Calcagni:2018lyd,Calcagni:2018gke,Calcagni:2018fid}, which recasts the original nonlocal equations for a single tensor field in a higher-dimensional localized form with a finite number of fields and their time derivatives. The purpose of this paper is to adapt the diffusion method to fractional QFTs, including gravity, but before doing that there is another issue that must be addressed: how to represent Lorentzian operators such as $\B^\g$?

The Riesz fractional Laplacian \cite{Rie49} is one of the first examples of a differential operator in $D$-dimensional fractional calculus \cite{MR,Pod99,SKM,KST} and a special case of a complex power of an elliptic operator \cite{Seeley:1967ea,Hor68,Hor71,Shu01}. However, its coordinate-dependent definition makes it unsuitable for Lorentz-invariant or manifestly covariant QFTs where the kinetic operator should be well-defined on any curved background and in any coordinate system. This is also the reason why the construction of QFTs with fractional derivatives is challenging \cite{Calcagni:2011kn,Calcagni:2011sz,Calcagni:2012qn,Calcagni:2012rm,Calcagni:2016azd,Calcagni:2018dhp,Calcagni:2021ljs,Calcagni:2021aap}. These theories are difficult because fractional derivatives and integrals are defined as integro-differential operators in spacetime coordinates which are not Lorentz invariant. Also, the quantum propagator from fractional derivatives is non-analytic due to its dependence on $|k^2|$ \cite{Calcagni:2021aap,Calcagni:2021ljs}; strictly speaking, analyticity is not necessary to formulate a well-defined QFT \cite{Anselmi:2025uzj} but it helps in its handling.

Because of all this, it is more convenient to turn from the theory of fractional powers of pseudo-differential operators to the theory of fractional powers of positive operators, summarized in the user-friendly textbooks \cite{Ama95,MaSa} (see also \cite{Hen81} for an older account). The mathematical literature on the subject has a long history. Fractional powers of a bounded operator $A$ were explored already in the first half of the XX century \cite{Hil39,Boc49}. Several definitions of $A^\a$ were proposed since the 1950s \cite{Phi52,Sol58,KrPu,KrSo,Phi59a,Phi59b,Bal59,Bal60,
	Kat60,Yos60,Kat61,Wat61,Yos61,Kom66,Kom67,Kom69a,Kom69b,Kom70,Wes74,MSM88,MS91} but the current and most studied definition of fractional powers $A^\a$ for a closed linear positive operator $A$ is due to Balakrishnan \cite{Bal60} and Komatsu \cite{Kom66,Kom67}, who also proved the composition rule $A^\a A^\b=A^{\a+\b}$. The Cauchy problem for these operators \cite{Ama95,MaSa} has been a constant preoccupation since the beginning \cite{Phi59a,Phi59b,Bal60}. In contrast, less attention has been given to self-adjoint extensions of $A^\a$ when $A$ is not self-adjoint \cite{Kat61}.

The d'Alembertian $\B$ in Lorentzian signature is not a positive operator and the above results have a limited application to the operator $(-\B)^\g$ in Euclidean spacetime. The scattering amplitudes of fractional QFT can be manipulated in Euclidean signature and then continued to Lorentzian signature with little or no trouble \cite{Calcagni:2022shb,Trinchero:2012zn,Trinchero:2017weu,Trinchero:2018gwe,Calcagni:2021ljs} but the problem is that the basic Lorentzian action with such operators is not Hermitian and, therefore, does not represent a classical dynamics. Thus, we need a different operator $\Bd^\g$ (``box dot gamma'') with the same anomalous scaling as $(-\B)^\g$ but giving rise to a Hermitian Lorentzian action and a unitary theory. Any representation of the operator $\Bd^\g$ should obey the following requirements:
\begin{enumerate}
	\item[(I)] To be well-defined for the non-positive operator $\B$.
	\item[(II)] To be self-adjoint, $(\Bd^\g)^\dagger=\Bd^\g$.
	\item[(III)] To be convenient to solve the Cauchy problem (problem of initial conditions).
	\item[(IV)] To yield a ghost-free spectrum.
\end{enumerate}
(I) The first constraint is to guarantee mathematical consistency. (II) Self-adjointness is necessary in order for the action to be Hermitian and for the theory to have a well-defined classical limit. Taking a scalar-field example, a non-self-adjoint kinetic operator $\phi\cK\phi$ would lead to a complex action $S$ even for real field configurations $\phi$ and, thus, to an erratic path integral of $\exp(\rmi S)\propto\exp(-\phi\Re\cK\phi)$ which may blow up in the classical limit or at saddle points. By the same token, time evolution may not conserve energy and can lead to a non-conservation of probability. (III) The third is a technical but fundamental condition in order to solve the classical equations of motion, a notoriously difficult problem in nonlocal theories. Certain definitions and representations of nonlocal operators can be mathematically impeccable but practically (for the physicist) unsuitable for this task. Finally, (IV) is a basic condition that, according to our current understanding, any viable quantum field theory should satisfy to be physical and stable.

In this paper, we construct an operator $\Bd^\g$ satisfying requirements (I)--(III) and briefly check property (IV), leaving a full analysis of (IV) to a separate publication \cite{CaBr}. In summary, self-adjointness is achieved by replacing the operator $(-\B)^\g$ with a different one given by the absolute value 
\be\label{eq1}
(-\B)^\g\to |\B|^\g\coloneqq (\B^2)^{\frac{\g}{2}}\,.
\ee
Despite the misleading appearance of the absolute value $|\B|$, we build one representation which allows for an \emph{analytic} treatment of this operator in momentum space and for a meaningful problem of initial conditions. An immediate physical consequence of the replacement \Eq{eq1} is the conversion of the particle modes associated with this operator to ghosts and, through a careful projection, their removal from the asymptotic states in scattering amplitudes. In other words, the physical spectrum of $(\B^2)^{\frac{\g}{2}}$ is empty and all the particle content in a fractional QFT is due to other factors in the kinetic terms, e.g., the first, standard factor in $\B+(\B^2)^{\frac{\g}{2}}$.

After reviewing the definition and representations of complex powers of positive operators in Section~\ref{sec2}, we construct several representations of the fractional d'Alem\-bertian in Section~\ref{sec3}. In particular, in Sections~\ref{sec3a} and \ref{sec3b} we recall the properties of the Lorentzian $\B$ operator and discuss why the known results of Section~\ref{sec2} do not immediately apply. In Sections~\ref{sec3c} and \ref{sec3d}, we offer two representations of the operator $(-\B)^\g$, while in Sections~\ref{sec3e} and \ref{sec3f} we build two representations of the self-adjoint operator $|\B|^\g$. The master representation $\Bd^\g$ we propose to use in fractional QFTs and quantum gravity is \Eqq{Bdot}. In Section~\ref{sec4}, we point out how the requirement of self-adjointness affects the spectrum of the theory and the way to eliminate ghost degrees of freedom. In Section~\ref{sec5}, we show that the domain of our fractional d'Alembertian can be restricted to the space of ultra-distributions and we compare it with the fractional operators defined on such space in the literature \cite{Barci:1996br,Barci:1998wp}. The novelties of our results with respect to the ultra-distributions approach are the proposal of a new fractional operator, an in-depth understanding of fractional d'Alembertians more centered on complex contours and spectral representations and, finally, a wider application to QFT. In particular, in Section~\ref{ultrun} we argue that ultra-distributions are possibly less convenient for a perturbative QFT. Section~\ref{sec6} is devoted to a novel application of the diffusion method to the Cauchy problem in fractional field theories. Taking a scalar field with a realistic ``$\B+\Bd^\g$'' kinetic term, in Section~\ref{inicon} we prove that the number of initial conditions to specify is two, since none is coming from $\Bd^\g$. We also generalize this conclusion to a much wider class of nonlocal theories that includes NLQG. Conclusions and a discussion on how to apply these results to quantum gravity are in Section \ref{concl}. 

The present work is written by physicists for physicists and does not pretend to cover all the mathematical details of operatorial representations in a rigorous way. Its novel results can still be of interest for mathematicians.


\section{Complex powers of positive operators}\label{sec2}

In this section, we review known results on complex powers of positive operators. Let us first recall some basic definitions. An operator $A$ is:
\begin{itemize}
	\item \emph{closed} if it is densely defined on its domain ${\rm D}(A)$ and its graph $\cG(A)=\{(f,g)\in {\rm D}(A)\times {\rm D}(A)\,:\,g=A[f]\}$ is closed in the product space therein, i.e., if the sequence $(f_n,g_n)\in\cG(A)$ converges to some $(f,g)\in{\rm D}(A)\times {\rm D}(A)$, then $(f,g)\in\cG(A)$;
	\item \emph{linear} if it obeys the additivity property $A[a\, f+ b\, g]=aA[f]+bA[g]$ for all $f,g\in{\rm D}(A)$ and the homogeneity property $A[a f]=a A[f]$ for any constant $a$ and any $f\in{\rm D}(A)$;
	\item \emph{positive} if it is linear on a Hilbert space $\cH$ with inner product $\langle\cdot,\cdot\rangle$ and $\langle A[f],f\rangle\geq 0$ for all $f\in\cH$. If $A$ is self-adjoint ($A^\dagger=A)$, then this is equivalent to state that the spectrum $\s(A)$ is non-negative.
\end{itemize}
The general definition of the complex power of a closed linear positive operator $A$ is given by the Dunford integral \cite{Ama95,Kom66}
\be\label{dunf}
A^{-\a}\coloneqq \frac{1}{2\pi\rmi}\int_\G\rmd z\,(-z)^{-\a}\,(z+A)^{-1}\,,\qquad \Re\,\a>0\,,
\ee
which is the operatorial counterpart of the Cauchy integral. Here $(z+A)^{-1}$ is the resolvent of $A$ and $\G$ is a contour encircling the spectrum $\s(A)$ counter-clockwise and avoiding the branch cut of the function $(-z)^{-\alpha}$  that we conventionally choose along the positive real axis. Rewriting the integral for $0<\Re\,\a<1$, one obtains the \emph{Balakrishnan representation} \cite{Ama95,MaSa,Bal60,Kom66,Kom67}
\be\label{bala}
A^{-\a}=\frac{\sin(\pi\a)}{\pi}\int_0^{+\infty}\rmd s\, s^{-\a}(s+A)^{-1}\,,\qquad 0<\Re\,\a<1\,.
\ee
Furthermore, if $-A$ is the infinitesimal generator of a strongly continuous semi-group $\{\rme^{-\t A}\,:\,\t\geq 0\}$ such that $\Vert\rme^{-\t A}\Vert\leq M\rme^{-\rho \t}$, where $M\geq 1$ and $\rho>0$ are finite, then \Eq{bala} is equivalent to the \emph{Balakrishnan--Komatsu representation} \cite{Ama95,MaSa,Kom66,Kom67}
\be\label{rep:exp}
A^{-\a}=\frac{1}{\G(\a)}\int_0^{+\infty}\rmd \t\, \t^{\a-1}\rme^{-\t A}\,,\qquad \Re\,\a>0\,,
\ee
where $\G$ is Euler's gamma function, the range of $\a$ is extended to all positive real values and we used the semi-group identity $(s+A)^{-1}=\int_0^{+\infty}\rmd \t\,\rme^{-\t(s+A)}$ and Fubini--Tonelli theorem \cite{Rud76}.

The above formul\ae\ can be extended to a positive exponent by the composition rule
\be
A^\g\coloneqq A^n A^{\g-n}\,,\qquad n-1<\Re\,\g<n\,,
\ee
where $0<\Re\,\a=\Re(n-\g)<1$. Also, if $0\in\s(A)$ all the above expressions are extended through the limit
\be\label{regA}
A^\g\coloneqq \lim_{\ve\to 0} (A+\ve)^\g\,.
\ee


\section{Complex powers of the d'Alembertian}\label{sec3}

To see whether and how the above representations can apply to the d'Alembertian $\B$, we must first check its properties. The desired fractional operator $\Bd^\g$ is presented in \Eqq{Bdot}.


\subsection{Properties of the d'Alembertian}\label{sec3a}

We denote by $C^\infty(\mathbb{R}^D)$ the space of smooth functions (derivable infinitely many times) on $\mathbb{R}^D$ and by $\cS(\mathbb{R}^D)\subset C^\infty(\mathbb{R}^D)$ the space of Schwartz functions, i.e., smooth functions which rapidly decay at $x^\mu\to\pm\infty$ with all their derivatives. The space of all distributions (generalized functions) is denoted by $\cD'(\mathbb{R}^D)$, a subset of which is the space $\cS'\subset \cD'$ of tempered distributions dual of $\cS$. $\cS$ is dense in the space of distributions, so that $\overline{\cS}=\cD'$ (where a bar denotes closure of the space). Examples: $\rme^{-x^2}\in \cS(\mathbb{R})$; $\de(x)\in\cS'(\mathbb{R})$; $\rme^{x^2}\in\cD'(\mathbb{R})\setminus\cS'(\mathbb{R})$; $\erf(x)\in \cD'(\mathbb{R})\setminus\cS(\mathbb{R})$. 

Other spaces of distributions $\subset\cD'$ were introduced in \cite[chapter IV]{GeS2}. In Section~\ref{uldi}, we will discuss the space of ultra-distributions $\zeta'$ \cite{Seb58,Has61,Morimoto:1975vi,Mor1,Mor2,Bollini:1994pg}, of special importance for fractional pseudo-differential operators. 

Having introduced various functional spaces whereon to consider the standard d'A\-lem\-bertian, we now turn to its properties of closeness, linearity, self-adjointness and real-valued spectrum.
\begin{enumerate}
	\item The operator $\B$ is closed in $\cS(\mathbb{R}^D)$. 
	\item The operator $\B$ is linear, since $\B[a\, f+ b\, g]=a\B f+b\B g$ for any covariantly constant $a,b$ and any $f,g\in{\rm D}(\B)$.
	\item The operator $\B$ is self-adjoint ($\B^\dagger=\B$) in the natural inner product of the Lebesgue integral on Schwartz functions, $\langle \B^\dagger f, g\rangle\coloneqq\langle f,\B g\rangle=\langle \B f, g\rangle$ for $f,g\in \cS(\mathbb{R}^D)$, which is obtained by dropping total derivatives after integration by parts.
	\item The spectrum of the operator $\B$ contains $0\in\s(\B)$. Hence, we use the regularization \Eq{regA} implicitly in all formul\ae, $-\B=\lim_{\ve\to 0}(-\B+\ve)$.
\end{enumerate}
These properties depend on the functional space constituting the domain ${\rm D}(\B)$ of the d'Alembertian and, in particular, on the boundary conditions established for the test functions. The standard in local QFT is to 
identify the domain of the $\B$ operator with the space of distributions ${\rm D}(\B)=\cD'(\mathbb{R}^D)$, so that the above checklist is satisfied in a weak sense (i.e., in the sense of distributions) because $\overline{\cS}=\cD'$. On the other hand:
\begin{enumerate}
	\item[5.] The operator $\B$ is \emph{not} positive since its spectrum $-k^2\in\mathbb{R}$ can take either sign in Lorentzian signature.
\end{enumerate}
The last property holds because $\B$ is a hyperbolic operator, i.e., an operator entering a hyperbolic differential equation. On Minkowski spacetime,
\be
\B=-\frac{\p^2}{\p x_0^2}+\sum_{i=1}^{D-1}\frac{\p^2}{\p x_i^2}\,.
\ee
This is just an example and all the representations below are valid for the gauge covariant d'Alembertian $\B=\N_\mu\N^\mu$ on any background.


\subsection{Balakrishnan--Komatsu representation of \texorpdfstring{$(-\B)^\g$}{Boxg}}\label{sec3b}

Property 5 is the key problem forbidding to apply the representations in Section~\ref{sec2} at face value. For example, in \cite{Calcagni:2021ljs,Calcagni:2021aap} the Balakrishnan--Komatsu representation \Eq{rep:exp} was used:
\be\label{Schw}
(-\B)^\g=\frac{1}{\Gamma(n-\g)}\int_0^{+\infty}\rmd\tau\,\tau^{n-\g-1}\,(-\B)^n\rme^{\t\B},\qquad n-1<\g<n\in\mathbb{N}\,.
\ee
This holds in Euclidean spacetime ($\B=\N^2$ is the $D$-dimensional Laplacian and $-\B$ is positive), since in momentum space
\be\label{Schw2}
(k^2)^\g=\frac{(k^2)^n}{\Gamma(n-\g)}\int_0^{+\infty}\rmd\tau\,\tau^{n-\g-1}\,\rme^{-\t k^2}
\ee
and the exponential always makes the integral converge. One can use \Eq{Schw2} also on functions in Lorentzian signature with positive eigenvalues $k^2>0$. For example, one can safely employ \Eq{Schw2} to derive the nonlocal equations of motion of a Lorentzian fractional QFT with the purpose of finding solutions with $k^2>0$. However, \Eq{Schw} is ill-defined in general in Lorentzian signature and we must look for alternatives. Even in the most optimistic case where one could make sense of a non-Hermitian fractional QFT with the $(-\B)^\g$ operator, for instance by inventing a PT-invariant formulation with viable properties \cite{El-Ganainy:2018ksn,Ashida:2020dkc}, the issue of the ill-defined representation \Eq{Schw} would persist.


\subsection{Balakrishnan representation of \texorpdfstring{$(-\B)^\g$}{Boxg}}\label{sec3c}

The problems with \Eq{Schw} suggest to take a step back and utilize \Eq{bala} instead of \Eq{rep:exp}, with $A=-\B$:
\be\label{bala2}
(-\B)^\g=\frac{\sin[\pi(n-\g)]}{\pi}\int_0^{+\infty}\rmd s\, s^{\g-n}(-\B)^n(s-\B)^{-1}\,.
\ee
Contrary to \Eq{Schw}, this expression converges for any sign of the spectrum of $\B$ and is therefore well-defined. However, it is not suitable to find the classical solutions of the theory. In fact, consider the toy model
\ben
\cL=\frac12 \phi(-\B)^\g\phi\,,
\een
with equation of motion $(-\B)^\g\phi=0$. Using the representation \Eq{bala2}, we have to solve $(-\B)^n(s-\B)^{-1}\phi=0$. In the presence of inverse powers of the d'Alembertian such as in $\cL=\phi\B^{-1}\phi/2$, a well-known trick is to make a nonlocal field redefinition $\chi\coloneqq \B^{-1}\phi+\la$, where $\la$ is an harmonic function ($\B\la=0$). Then, $\cL=\chi\B\chi/2$ up to a total derivative. In our case, we have an extra $s$ dependence and we have to modify the field redefinition as $\chi(s,x)\coloneqq (s-\B)^{-1}\phi(x)+\la_s(x)$, where $\la_s$ is an eigenfunction of $\B$ with eigenvalue $s$ ($(s-\B)\la_s=0$). Then, $\phi=(s-\B)\chi$ and (up to total derivatives)
\ban
\cL&=&\frac12 \phi(-\B)^\g\phi=\frac{\sin[\pi(n-\g)]}{2\pi}\int_0^{+\infty}\rmd s\, s^{\g-n}\phi(-\B)^n(s-\B)^{-1}\phi\\
&=&\frac{\sin[\pi(n-\g)]}{2\pi}\int_0^{+\infty}\rmd s\, s^{\g-n}[(-\B)^n\chi](s-\B)\chi\,,
\ean
which is a higher-order local Lagrangian requiring a finite number of initial conditions. However, the above integral does not converge and the field redefinition does not work as hoped.


\subsection{Logarithmic representation of \texorpdfstring{$(-\B)^\g$}{Boxg}}\label{sec3d}

A natural definition of complex powers of complex numbers expresses the power law as the exponential of a logarithm. We explored this possiblity in the quest for a viable representation of the fractional d'Alembertian. Instating a length scale $\lst$ to make the operator dimensionless,
\be
(-\lst^2\B)^\g\coloneqq \rme^{\g\ln (-\lst^2\B)}\coloneqq \sum_{n=0}^{+\infty} \frac{\g^n}{n!}[\ln (-\lst^2\B)]^n\,.
\ee
Using the representation of the logarithm
\be\label{loga}
\ln\left(-\lst^2\B\right)\coloneqq\int_0^{+\infty}\rmd s\left(\frac{1}{s+\lst^{-2}}-\frac{1}{s-\B}\right),
\ee
we can rewrite $(-\B)^\g$ as an infinite series of $(s-\B)^{-1}$ operators, which is no more convenient than \Eq{bala2} and, therefore, we do not use in what follows.


\subsection{Self-adjoint Fresnel representation of \texorpdfstring{$|\B|^\g$}{|Box|g}}\label{sec3e}

Insisting on raising the operator $\B$ to a fractional power inevitably leads to the problems of non-positivity and non-self-adjointness. Let us therefore turn our attention to the operator $|\B|^\g$ and its representation
\be
\B_{\rm F}^\g\coloneqq |\B|^n |\B|^{-(n-\g)},\qquad n-1<\Re\,\g<n\in\mathbb{N}\,,
\ee
where we use the Fresnel (F) integral (see formula 3.763.9 of \cite{GR})
\be\label{cos}
\B_{\rm F}^\g=\frac{1}{\G(n-\g)\,\cos\frac{\pi(n-\g)}{2}}\int_0^{+\infty}\rmd\tau\,\tau^{n-\g-1}|\B|^n\,\cos(\t\B)\,.
\ee
The idea behind this formula (see \cite[Section~4.1.7]{Calcagni:2021ljs}) is to apply the Balakrishnan--Komatsu representation \Eq{rep:exp} to the operators $\rmi\B$ and $(\rmi\B)^\dagger$ and then to take the average. The divergent exponential in \Eq{Schw} is changed into a ``phase" $\exp(\rmi\B)$ and the integral becomes of Fresnel type. The average with its adjoint $\exp(-\rmi\B)$ gives the self-adjoint expression \Eq{cos}. More simply, one takes the real part of \Eq{Schw}.

Equation \Eq{cos} may be suitable to solve the problem of initial conditions. Just like in the representations in Sections~\ref{sec3c} and \ref{sec3d}, we have recast the nonlocal operator ``$\B^\g$" as a parametric integral of another nonlocal operator. Trading one type of nonlocality with another can be advantageous only if we know how to manipulate the end point. The exponential $\exp(\t\B)$ in \Eq{Schw} can be treated with the diffusion method \cite{Calcagni:2018lyd} but \Eq{Schw} is ill-defined. \emph{Vice versa}, \Eq{cos} is well-defined and it is possible to treat the nonlocal operator $\cos(\t\B)$ with the diffusion method. We further explore this in Section~\ref{sec6}, where, however, we mainly use the following simpler operator.


\subsection{Self-adjoint Balakrishnan--Komatsu representation of \texorpdfstring{$|\B|^\g$}{|Box|g}}\label{sec3f}

We now consider the very simple alternative of raising the squared d'Alembertian $\B^2$ to a fractional power. The Lorentzian operator $\B^2$ has ${\rm D}(\B^2)=\cD'(\mathbb{R}^D)$ and is closed, linear, positive (with spectrum $(k^2)^2\geq 0$) and self-adjoint ($\B^{2\dagger}=\B^\dagger\B^\dagger =\B\B=\B^2$) in $\cD'(\mathbb{R}^D)$. Therefore, we can replace our starting point $(-\B)^\g$ with a different operator with different physical properties,
\be\label{newbox}
(-\B)^\g\to |\B|^\g \coloneqq (\B^2)^{\frac{\g}{2}}
\ee
and apply the Balakrishnan--Komatsu representation \Eq{rep:exp} to $\B^2$ for a power $\a=\g/2-n$ to obtain the ``box dot gamma'' self-adjoint operator:
\be
\boxd{\begin{matrix*}[l]
		\Bd^\g &\coloneqq (\B^2)^n (\B^2)^{\frac{\g}{2}-n}\vphantom{\frac{1}{(\B^2)^{\frac{\g}{2}-n}}}\\
		&=\dfrac{1}{\Gamma(n-\g/2)}\displaystyle\int_0^{+\infty}\rmd\tau\,\tau^{n-\frac{\g}{2}-1}\,(\B^2)^n\rme^{-\t\B^2},\qquad n-1<\Re\,\frac{\g}{2}<n\in\mathbb{N}\,,
\end{matrix*}}\label{Bdot}
\ee
where $\tau\geq 0$ is a non-negative integration parameter. When $\g=1$, one gets the absolute value of the ordinary d'Alembertian, since for a real spectrum $\sqrt{\B^2}=|\B|$. Also, the new operator inherits all the properties of the Balakrishnan representation including linearity and the composition rule for powers. Finally, its representation in momentum space (Fourier transform $\widetilde{\Bd^\g}(k^2) \tilde f(k)$) is analytic because we do not have absolute values around such as $|k^2|$ (however, its Green's function will happen to be non-analytic). In summary:
\begin{itemize}
	\item $\Bd^\g[a\, f+ b\, g]= a\Bd^\g f+ b \Bd^\g g$.
	\item $(\Bd^\g)^\dagger=\Bd^\g$.
	\item $\Bd^n=|\B|^n$, $n\in\mathbb{N}$.
	\item $\Bd^\g \Bd^\a=\Bd^{\g+\a}$.
	\item $\widetilde{\Bd^\g}(z)$ is analytic in $z$.
\end{itemize}
A proof of the composition rule can be found along the same steps detailed in \cite{Ama95,MaSa}. To the above properties, we should add the Leibniz rule for the action of the fractional operator on the product of functions. The expression for $\Bd^\g[fg]$ is given in eq.~(B.11) of \cite{Calcagni:2022shb}, which is valid for any representation of the operator ``$\B^\g$.''

An advantage of \Eq{Bdot} with respect to \Eq{cos} is that we can adapt the diffusion method more naturally to the operator $\exp(-\t\B^2)$, as we show in Section~\ref{sec6}. 


\section{Spectrum of the purely fractional operator}\label{sec4}

Having \Eq{Bdot} as a candidate for the representation with best properties for a physical fractional theory of field, let us see how the spectrum changes when taking $(\B^2)^{\g/2}$ instead of $(-\B)^\g$. A very convenient tool to study the spectrum of an operator is the K\"allén--Lehmann representation of the propagator. This representation highlights the particle modes that can go on-shell as asymptotic states in scattering amplitudes at any given perturbative order, while differentiating them from off-shell states that do not make the propagator diverge. Moreover, the behaviour and sign of the spectral density functions associated to each mode tell whether on-shell ghost states are present or not.

In this section, we work on Minkowski spacetime and, for simplicity, we consider the (unphysical) case of a purely fractional kinetic term. The full case $\B+(\B^2)^{\g/2}$ with the infrared term $\B$ will be studied elsewhere \cite{CaBr}. In the following, we take the physical case $\Im\,\g=0$ and ignore the mass, which amounts to a shift of $k^2$ \cite{CaBr}.


\subsection{Propagator of \texorpdfstring{$(-\B)^\g$}{(-Box)g}}

Given a propagator\footnote{Following the majority of QFT textbooks, we take ``propagator'' and ``Green's function'' as synonyms, both with and without the $\rmi$ factor. Sometimes one differentiates between the Green's function $1/(k^2+m^2)$ and the propagator $\rmi/(k^2+m^2)$ but we will not do it here.} $\tilde G(-k^2)$ in momentum space, consider its K\"allén--Lehmann representation starting from the Cauchy integral
\be\label{Gz}
\tilde G(-k^2)=\frac{1}{2\pi\rmi}\int_{\G}\rmd z\,\frac{\tilde G(z)}{z+k^2-\rmi\e}\,,
\ee
where we took the Feynman prescription $k^2\to k^2-\rmi\e$ and $\G$ is a counter-clockwise contour encircling the pole $z=-k^2$ and such that $\tilde G(z)$ is holomorphic inside and on $\G$. 

\begin{figure}[ht]
	\bc
	\includegraphics[width=10cm]{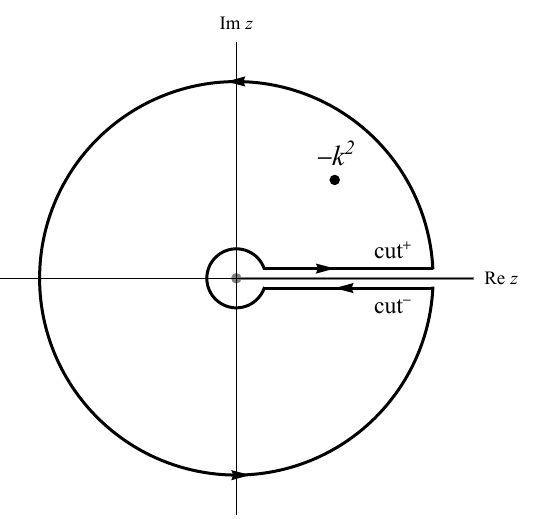}
	\ec
	\caption{\label{fig1} Maximal contour $\G$ of the Cauchy representation of the propagator \Eq{Gz1} of the operator $(-\B)^\g$. The contour encompasses the pole at $z=-k^2$. The half-lines ${\rm cut}^\pm$ run along the branch cut at $\Im\,z=0$, $\Re\,z>0$, which starts at the branch point $z=0$.}
\end{figure}

In the case of the operator $(-\B)^\g$, the propagator in momentum space and in the complex $z$-plane is
\be\label{Gz1}
\tilde G(-k^2)=\frac{1}{(k^2)^\g}\,,\qquad \tilde G(z)=(-z)^{-\g}\,,
\ee
integrated on the contour shown in Fig.~\ref{fig1} (maximal extension of the contour $\G$ in \Eq{Gz}). This contour avoids the branch point $z=0$ and the branch cut at $\Re\,z>0$. The corresponding K\"allén--Lehmann representation is \cite{Calcagni:2021ljs,Calcagni:2022shb}
\be\label{propold}
\tilde G(-k^2)=\int_0^{+\infty}\rmd s\,\frac{\rho(s)}{s+k^2-\rmi\e}\,,\qquad \rho(s)=\frac{\sin(\pi\g)}{\pi}\frac{1}{s^\g}\,.
\ee


\subsection{Spectrum of \texorpdfstring{$(-\B)^\g$}{(-Box)g}}

The spectrum consists of a continuum of massive particles of squared mass $s\geq 0$ and a spectral density $\rho(s)$. These particles can appear as intermediate states in perturbation theory, according to the optical theorem (\Eq{propold} has a non-vanishing imaginary part). In order to avoid ghosts, it must be $\rho(s)>0$ for all $s\geq 0$, which happens if $0<\g<1$, $2<\g<3$ and so on (we exclude integer values). In alternative, since the mass $s$ in the propagator $1/(s+k^2)$ is non-negative, one can define the propagator and the amplitudes with the Anselmi--Piva prescription  \cite{Anselmi:2017yux,Anselmi:2017lia,Anselmi:2018bra,Anselmi:2019rxg,Anselmi:2021hab,Anselmi:2022toe,Anselmi:2025uzj} and make all these modes purely virtual (fake particles or \emph{fakeons}) order by order in perturbation theory. In this case, the spectrum of the theory would be empty and there would be no constraint on the values of $\g$, since any ghost mode in the spectral density $\rho(s)$ would be fakeonized. 

The spectrum of the operator $(\B^2)^{\g/2}$ is not the same or, more precisely, it must be fakeonized necessarily. This is a rather tricky point and it is worth dwelling on it in detail.


\subsection{Propagator of \texorpdfstring{$(\B^2)^\frac{\g}{2}$}{(Box2)g/2}}

The purely fractional propagator in momentum space is
\be\label{Gk2}
\tilde G(-k^2)=\frac{1}{[(k^2)^2]^\frac{\g}{2}}\eqqcolon\frac{1}{(k^4)^\frac{\g}{2}}\,.
\ee
To write the correct Cauchy representation, we have to ensure that, on one hand, the complex function $\tilde G(z)$ is well-defined and, on the other hand, that the contour $\G$ is such that the final result $\tilde G(-k^2)=\tilde G(k^2)$ is even in $k^2$. However, it is not difficult to show (\ref{appA}) that the causal propagator \Eq{Gz} can be evaluated consistently only on the contour in Fig.~\ref{fig1}. This eventually leads to the same complex function $\tilde G(z)$ in \Eq{Gz1} and to the same spectral representation \Eq{propold}, which is not invariant under the reflection $k^2\to -k^2$. Therefore, \Eq{propold} is not a representation of the propagator \Eq{Gk2}.

The origin of this problem is that \Eq{Gk2} cannot be defined analytically by \Eq{Gz} and this expression must be replaced by a non-analytic definition. Intuitively, this happens because we are trying to write $|\B|^{-\g}$ in momentum space. The solution is to find a generalized K\"allén--Lehmann representation symmetric under $k^2\to -k^2$; it turns out that this representation entails some non-analytic steps.

\begin{figure}[ht]
	\bc
	\includegraphics[width=10cm]{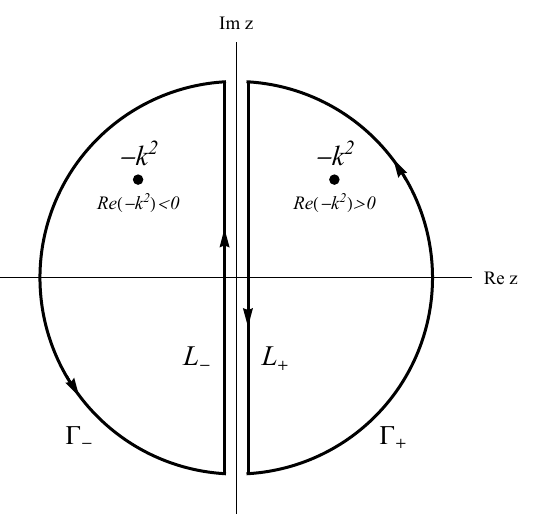}
	\ec
	\caption{\label{fig2} Contour $\G=\G_+\cup\G_-$ of the Cauchy representation of the propagator \Eq{Gtota}. $\G_+$ and $\G_-$ are mutually disjoint pieces covering, respectively, the $\Re\,z>0$ and the $\Re\,z<0$ half-plane making up definition \Eq{Gtota}. The vertical lines $L_\pm$ run along the discontinuity at $\Re\,z=0$.}
\end{figure}

To determine the contour $\G$ associated with a self-adjoint operator, we note that $\tilde G(z)=(z^2)^{-\g/2}$ as a function of $z^2$ has a discontinuity at $\Re\,z^2<0$ and $\Im\,z^2=0$, i.e., on the whole imaginary axis $\Re\,z=0$ in the $z$-plane. Therefore, we divide $\G$ into two disjoint counter-clockwise contours $\G_+$ and $\G_-$ in, respectively, the half-plane $\Re\,z>0$ and $\Re\,z<0$. $\G_+$ and $\G_-$ are made of a semi-circle at infinity and a line running parallel to the imaginary axis (Fig.~\ref{fig2}).

Depending on the sign of $-k^2$, one of the two contours must give zero, so that \Eq{Gz} becomes
\bs\label{Gtot}\ba
\tilde G(-k^2)&=& \Theta[\Re(-k^2)]\,\tilde G_+(-k^2)+\Theta[\Re(k^2)]\,\tilde G_-(-k^2)\,,\label{Gtota}\\
\tilde G_\pm(-k^2) &=& \frac{1}{2\pi\rmi}\int_{\G_\pm}\rmd z\,\frac{\tilde G(z)}{z+k^2}\,,\qquad \tilde G(z)=(z^2)^{-\frac{\g}{2}}\,,\label{Gz2}
\ea\es
where $\Theta$ is Heaviside step function. The operation leading to \Eq{Gtot} is not analytic. It is tantamount to write $(z^2)^{-\g/2}=(\sqrt{z^2})^{-\g}$, take the two branches $\sqrt{z^2}=\pm z$ on the $\Re\,z\gtrless 0$ half-plane, respectively, and glue them together.

A double-checked brute-force calculation of \Eq{Gtot} on the $z$-plane is reported in \ref{appB} and yields
\be\label{KLfin}
\boxd{\tilde G(-k^2)=\int_0^{+\infty}\rmd s\,\frac{\rho(s)}{s+k^4}\,,\qquad \rho(s)=\frac{\sin\frac{\pi\g}{2}}{\pi}\frac{1}{s^\frac{\g}{2}}\,.}
\ee
A second, much faster way to get the same result is to consider the Cauchy representation of $\tilde G(-k^2)$ in the complex plane of $w=-z^2$:
\be\label{Gw}
\tilde G(-k^2)=\frac{1}{2\pi\rmi}\int_{\G_w}\rmd w\,\frac{\tilde G(w)}{w+k^4}\,,\qquad \tilde G(w)=(-w)^{-\frac{\g}{2}}\,,
\ee
where $\G_w$ is a counter-clockwise contour encircling the pole $w=-k^4$ and such that $\tilde G(w)$ is holomorphic inside and on $\G_w$. In practice, $\G_w$ is the contour shown in Fig.~\ref{fig1} but in the $w$-plane and with pole $-k^4$ instead of $-k^2$. This propagator is exactly of the same form studied in \cite{Calcagni:2021ljs,Calcagni:2022shb} and is much easier to compute than \Eq{Gz}. Indeed, \Eq{KLfin} is nothing but 
\Eq{propold} with $\e=0$, $\g\to\g/2$ and $k^2\to k^4$.


\subsection{Spectrum of \texorpdfstring{$(\B^2)^\frac{\g}{2}$}{(Box2)g/2}}

While \Eq{propold} has a continuum of normal particles, \Eq{KLfin} has a continuum of complex-conjugate pairs, which are off-shell at the tree level but can appear as intermediate states in the optical theorem at higher loop orders. However, complex-conjugate pairs contain a ghost mode. Indeed, we can decompose the propagator of a pair with imaginary squared masses $\pm\rmi M^2$ as
\be
\frac{1}{k^4+M^4}=\frac{1}{2\rmi M^2}\left(\frac{1}{k^2-\rmi M^2}-\frac{1}{k^2+\rmi M^2}\right),
\ee
which has two modes with different residue sign. The ghost must be projected out of the spectrum of asymptotic states but doing so would leave the other particle, hence giving rise to a complex-valued spectrum. Therefore, complex-conjugate pairs should be projected out as a whole and this is done consistently with Lorentz invariance and at all orders in perturbation theory by the Anselmi--Piva prescription; we refer to \cite{Anselmi:2025uzj} for all the details of the procedure. With this prescription, the physical spectrum of a theory with purely fractional kinetic term $\phi\,(\B^2)^{\g/2}\phi$ is empty. In this sense, the purely fractional model is the analogue of the $p$-adic string with pole-less kinetic term $\phi\,\rme^{-\B}\phi$ \cite{Moeller:2002vx,Brekke:1987ptq}. An important difference, however, is that in the fractional case studied here the spectrum is empty because all degrees of freedom are fakeonic and they are projected out of the physical spectrum, while in the $p$-adic case the kinetic term is entire and there are no degrees of freedom at all.

Note that there is no restriction on the allowed values of $\g>0$ in \Eq{Gk2} apart from positivity, since the Anselmi--Piva prescription removes the pairs regardless of the sign of the spectral density $\rho$. Therefore, fractional QFTs featuring the operator $(\B^2)^{\g/2}$ can be unitary for values of $\g$ forbidden in the case of $(-\B)^\g$. These results have implications for FQG \cite{Calcagni:2021aap,Calcagni:2022shb} which will be discussed in greater detail in a separate analysis \cite{CaBr}. Briefly, the version of the theory with $(-\B)^\g$ \cite{Calcagni:2021aap,Calcagni:2022shb} has a continuum of modes that alter the standard graviton spectrum but suffers from the problems presented in the introduction. The replacement \Eq{eq1} makes the graviton the only particle in the spectrum and there is no continuum of ``fractional'' modes among the asymptotic states of scattering amplitudes. This provides a remarkable simplification of the physical interpretation of that theory but it also bars the possibility to verify fractional QFTs via the measurement of cross-sections alone. The constraints found in \cite{Calcagni:2022shb} on the range of $\g$ are also lifted.


\section{Comparison with the ultra-distributions approach}\label{sec5}

In Section \ref{sec3}, we have maintained the usual space of distributions $\cD'(\mathbb{R}^D)$ and constructed self-adjoint representations of the fractional d'Alembertian thereon, which we applied in Section \ref{sec4} to study the spectrum of $|\B|^\g$ in a QFT context. An alternative approach is to select the more specialized functional space $\zeta'$ of ultra-distributions \cite{Seb58,Has61,Morimoto:1975vi,Mor1,Mor2,Bollini:1994pg} and work always on Minkowski spacetime and, almost exclusively, in momentum space. Here we compare this approach with ours, both mathematically and physically, and find that our approach may be regarded as more general, since our representations are valid on the larger functional space $\cD'$ and the operator $\B$ is on any metric background. The reader uninterested in this comparison can skip this section. 


\subsection{Fractional propagators in momentum space}

One of the first to use a Lorentzian fractional d'Alembertian in QFT were Bollini, Giambiagi and Gonz\'ales in 1964 as a mathematical means to regularize scattering amplitudes in electrodynamics \cite{BGG}. The operator and its propagator were represented both in position and momentum space. The causal propagator was $(m^2+k^2-\rmi\e)^{-\a}$ for a massive field. Almost 30 years later, Marino considered the operator $(-\B)^{1/2}\to(k^2)^{1/2}$ in a model of electrodynamics, passing from Lorentzian to Euclidean signature and back when needed \cite{Mar91,Marino:1992xi}.

The problem of when the propagator respects Huygens' principle (i.e., when its support is on the light cone and does not spill over) was examined for the Lorentzian operators $(-\B)^\a$ \cite{Gia91,BGO,BG} and $\B^\a$
\cite{doA92}. In a massless scalar field model, expressions of four different propagators in position and momentum space were given starting from a definition of the fractional d'Alembertian in momentum space \cite{BG,doA92}. For $(-\B)^\a$, these propagators are: causal $(k^2-\rmi\e)^{-\a}$, anti-causal $(k^2+\rmi\e)^{-\a}$ (the Wightman function being the difference between causal and anti-causal), retarded $[-(k^0+\rmi\e)^2+\bm{k}^2])^{-\a}$ and advanced $[-(k^0-\rmi\e)^2+\bm{k}^2])^{-\a}$, with corresponding expressions in position space which were studied originally by Riesz \cite{Rie49}. The canonical quantization of a scalar field with fractional d'Alembertian was considered in \cite{doA92}, later generalized to generic form factors and extended to the case of an Abelian gauge field \cite{Barci:1995ad}.

The Fourier transform of the fractional d'Alembertian seems the most natural way to represent it, as done in the above-mentioned papers and in Section~\ref{sec4}. However, the functional spaces to which this operator belongs and on which it acts should be defined with care. These spaces are sub-spaces of $\cD'$.


\subsection{Ultra-distributions}\label{uldi}

Let us introduce the space $\tilde\zeta=\tilde\zeta(\mathbb{C}\times\mathbb{R}^{D-1})$ of the so-called ultra-analytic functions, entire holomorphic functions $\tilde\vp(k)=\tilde\vp(k^0,\bm{k})$ in the complex $k^0$-plane which are rapidly decreasing on any horizontal line $\Im\,k^0={\rm const}$. In the following, spatial directions play a passive role. The Fourier anti-transform 
\be\label{foua}
\vp(x)=\vp(x^0,\bm{x})=\cF^{-1}[\tilde\vp]=\int_{-\infty}^{+\infty}\frac{\rmd^Dk}{(2\pi)^{D}}\,\rme^{\rmi k\cdot x}\tilde\vp(k)
\ee
of an ultra-analytic function is an element of the space $\zeta\subset C^\infty$ of functions such that $|\p_0^n\vp(x)|\leq \rme^{-a|x^0|}$ for any $a,n>0$ and $x^0\to\pm\infty$. 

\begin{figure}[ht]
	\bc
	\includegraphics[width=13cm]{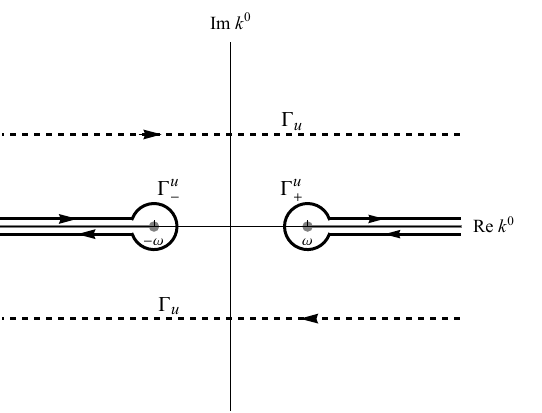}
	\ec
	\caption{\label{fig3} Path $\G_{\rm u}$ (dashed lines) in the $(\Re\, k^0,\Im\, k^0)$ plane and its deformation $\Gamma_+^{\rm u}\cup\Gamma_-^{\rm u}$ (solid thick curves) around the branch cuts $k^0\leq -\om$ and $k^0\geq \om$ (gray thick lines) for a massive field. Credit: adaptation of \cite{Calcagni:2021ljs}.}
\end{figure}

The dual of $\tilde\zeta$ is the space $\tilde\zeta'$ of linear distributions $\tilde\Psi$ represented by analytic functions $\tilde\psi(k)=\tilde\psi(k^0,\bm{k})$ and acting on $\tilde\zeta$ via the scalar product
\be\label{psivp}
\tilde\Psi[\tilde\vp]=(\tilde\psi,\tilde\vp)=\int_{\G_{\rm u}\times \mathbb{R}^{D-1}}\frac{\rmd^D k}{(2\pi)^D}\,\tilde\psi(k)\,\tilde\vp(k)\,,\qquad \tilde\psi(k)\in\tilde\zeta'\,,
\ee
where $\tilde\psi(k)$ is holomorphic in the domain $\cC_n=\{k^0:|\Im k^0|>n\}$ and $(k^0)^{-n}\tilde\psi(k)$ is bounded continuous in the domain $\cC_n^==\{k^0:|\Im\, k^0|\geq n\}$. $\G_{\rm u}$ (depicted in \cite{Calcagni:2021ljs,Belenchia:2014fda} and in Fig.~\ref{fig3}) runs from $-\infty$ to $+\infty$ along $\Im\, k^0>n$ and from $+\infty$ to $-\infty$ along $\Im\, k^0<-n$ for some $n\in\mathbb{N}$. If, furthermore, $\tilde\psi(k)$ is holomorphic for $\Im\, k^0\neq 0$ and has two branch cuts for $|k^0|>\om$ where $\om=\sqrt{|\bm{k}|^2+m^2}$, then $\G_{\rm u}$ can be deformed as in Fig.~\ref{fig3}.

Recall from Section~\ref{sec2} that the space $\cS'$ of tempered distributions is a subset of $\cD'$. Although also $\tilde\zeta'\subset\cD'$, it contains elements not in $\cS'$. An example of functional in $\tilde\zeta'\setminus\cS'$ is $\tilde\psi(k)=\rme^{\rmi k\cdot x}$. When $\Im\, k^0>0$, this blows up as $\rme^{(\Im\,k^0)x^0}$. A Fourier transform on a real function (\Eqq{psivp} with $\G=\mathbb{R}$) would not converge in general but \Eq{psivp} with $\G$ specified as above does. In this sense, \Eq{psivp} is the natural generalization of the Fourier transform on $\mathbb{R}^D$ in QFT. More precisely, the Fourier anti-transform of $\tilde\psi(k)$ is defined as $\Psi[\vp]=\cF^{-1}\tilde\Psi[\cF^{-1}\tilde\vp]=\int_{-\infty}^{+\infty}\rmd^Dx\,\psi(x)\,\vp(x)$, where
\be\label{soluz0}
\psi(x)=\int_{\G\times \mathbb{R}^{D-1}}\frac{\rmd^Dk}{(2\pi)^D}\,\rme^{\rmi k\cdot x}\,\tilde\psi(k)\in\zeta'\,.
\ee
The elements of $\tilde\zeta'$ and $\zeta'$ are called ultra-distributions. The ``unity'' ultra-distributions for the above choice of path $\G$ are $\tilde\psi(k)={\rm sgn}(\Im\,k^0)/2\in\tilde\zeta'$ and $\psi(x)=\de^D(x)\in\zeta'$ \cite{Barci:1996br,Barci:1998wp}.


\subsection{Fractional d'Alembertian and ultra-distributions}

The spaces $\tilde\zeta'$ and $\zeta'$ were designed to handle distributions with moderate singularities, such as those produced by fractional pseudo-differential operators \cite{Barci:1996br,Barci:1998wp}. Fractional operators $\cK(\B)$ have domain ${\rm D}[\cK(\B)]=\zeta'$ and propagators in position space belong to $\zeta'$. If the representation $\cK(-k^2)=\cK(k^0,\bm{k})$ in momentum space of $\cK(\B)$ is holomorphic on the quotient domain $\cC_n/\cC_p$ for some $n,p\in\mathbb{N}$ and such that $(k^0)^{-p}\cK(k^0,\bm{k})$ is bounded continuous in $\cC_n^=/\cC_p^=$, 
\be
\left|(k^0)^{-p}\cK(k^0,\bm{k})\right|\stackrel{|k^0|\to\infty}{\longrightarrow} c\in\mathbb{R}\,,
\ee
then $\cK(\B)$ applied to \Eq{soluz0} yields a well-defined expression \Eq{psivp} with $\tilde\psi(k)=\rme^{\rmi k\cdot x}\cK(-k^2)\in\tilde\zeta'$ and $\G$ running from $-\infty$ to $+\infty$ for $\Im\, k^0>n+p$ and from $+\infty$ to $-\infty$ for $\Im\, k^0<-(n+p)$. In particular, $\cK(-k^2)=(k^2)^\g$ is holomorphic on $\cC_n/\cC_p$ and bounded continuous in $\cC_n^=/\cC_p^=$ for any $n\geq 1$ and $\lceil 2\g\rceil\leq p<n$. Therefore, the fractional d'Alembertian $\cK(\B)=(-\B)^\g$ acts on ultra-distributions and one can restrict its domain to ${\rm D}[(-\B)^\g]=\zeta'$. 

While \cite{BGG,Mar91,Marino:1992xi,Gia91,BGO,BG,doA92} considered the fractional d'Alembertian and its propagators taking their naive Fourier transform, starting from the mid-1990s the construction of fractional operators on $\zeta'$ was made explicit in scalar and gauge QFT as a proof of concept \cite{Barci:1995ad,Barci:1996br,Barci:1998wp}, in causal sets \cite{Belenchia:2014fda} and in fractional QFTs \cite{Calcagni:2021ljs}.

After this review of the theory of ultra-distributions \cite{Bollini:1994pg} and its application to fractional d'Alembertians, we are ready to complement the discussion in Section~\ref{sec3a} and show that:
\begin{theo}
	The Balakrishnan--Komatsu representation \Eq{Bdot} of $|\B|^\g$ holds on the space of ultra-distributions and one can restrict its domain on ${\rm D}(\Bd^\g)=\zeta'$.
\end{theo}
\begin{proof}
First, we show that $\B$ is closed in $\zeta'$ and, from there, that also $\B^2$ is closed in $\zeta'$. In fact:
\begin{itemize}
	\item The operator $\B$ is closed in $\zeta'$. In fact, $\cK(-k^2)=-k^2$ is holomorphic on $\cC_n/\cC_p$ and bounded continuous in $\cC_n^=/\cC_p^=$ for any $2\leq p\leq n$, so that both $-k^2\in\tilde\zeta'$ and $-k^2\tilde\psi(k)\in\tilde\zeta'$ for any $\tilde\psi(k)\in\tilde\zeta'$. Then, it is easy to show that $-k^2\,:\,\tilde\zeta'\to\tilde\zeta'$ is dense in $\tilde\zeta'$ and its graph $\cG$ is closed. Then, $-k^2$ is closed in $\tilde\zeta'$ and so is the $\B$ operator in the space $\zeta'$.
	\item The operator $\B^2$ is closed in $\zeta'$. In fact, $\cK(-k^2)=(k^2)^2$ is holomorphic on $\cC_n/\cC_p$ and bounded continuous in $\cC_n^=/\cC_p^=$ for any $4\leq p\leq n$ and, moreover, the $\B$ operator is closed in $\zeta'$; the rest of the proof is as above.
\end{itemize}
Since $\B^2$ is linear, positive and closed in $\zeta'$, we conclude that the Balakrishnan--Komatsu representation \Eq{Bdot} is well-defined on the space $\zeta'$. Therefore, we can restrict the domain of $\Bd^\g$ to $\zeta'$.
\end{proof}

Since ${\rm D}(\Bd^\g)\supseteq\zeta'$, we can compare its propagator \Eq{KLfin} directly with the propagators found in the theory of ultra-distributions \cite{Barci:1996br,Barci:1998wp}. Apart from the above-mentioned causal, anti-causal, retarded, advanced and Wightman propagators, a fifth one is the principal-value (PV) propagator \cite{Barci:1996br} (see also \cite{Calcagni:2021ljs})
\be\label{pvG}
\tilde G_{\rm PV}(-k^2)=\frac12\left[(k^2-\rmi\e)^{-\g}+(k^2+\rmi\e)^{-\g}\right]= [\Theta(-k^2)+\cos(\pi\g)\,\Theta(k^2)]|k^2|^{-\g}\,,
\ee
which differs by a distributional factor with respect to \Eq{Gk2}. While \Eq{pvG} corresponds to the average of the propagators of the operators $(\B+\rmi\ve)^\g$ and $(\B-\rmi\ve)^\g$, in the operator \Eq{Bdot} these conjugate terms appear in a product,
\be
|\B|^\g =\lim_{\ve\to 0} (\B^2+\ve^2)^\frac{\g}{2}=\lim_{\ve\to 0}(\B+\rmi\ve)^\frac{\g}{2}(\B-\rmi\ve)^\frac{\g}{2}\,,
\ee
which leads to \Eq{Gk2} and \Eq{KLfin}. Therefore, \Eq{pvG} and \Eq{Gk2} are slightly different but both correspond to self-adjoint operators and both have an empty spectrum. In fact, \Eq{pvG} is nothing but the non-analytic tree-level propagator in the Anselmi--Piva prescription, obtained as the average analytic continuation of the Euclidean propagators. Consistently with the fact that particles prescribed with the Anselmi--Piva procedure are projected out of the physical spectrum of the theory, at the tree level the only asymptotic state associated with \Eq{pvG} is the vacuum \cite{Barci:1996ny}. This is in complete agreement with our findings in Section~\ref{sec4} that the spectrum of $|\B|^\g$ is empty.

A general conclusion we can draw is that, at the tree level and if defined as a self-adjoint operator, the fractional d'Alembertian does not carry new physical degrees of freedom. Unphysical modes corresponding to a non-Hermitian action or to complex-conjugate pairs, one of whose members is a ghost, are removed by a certain choice of contour in the complex plane of momenta.

The discrepancy between \Eq{Gk2} and \Eq{pvG} is not due to a different choice of contour. A direct comparison of the contours is not straightforward because the method to represent the propagator is also different, since our analysis is based on the very convenient complex $k^2$-plane \cite{Calcagni:2021ljs,Calcagni:2022shb,Briscese:2024tvc} while that of \cite{Barci:1996br,Barci:1998wp} is based on the $k^0$ complex plane. On general grounds, the relation between the two contours is not analytic and is made explicit in \ref{appC}. There, we show that the propagator $(k^4)^{-\g/2}$ is defined on a path in the complex $k^0$-plane that can be deformed homotopically to the path $\G_+^u\cup\G_-^u$ of Fig.~\ref{fig3}. From this, we conclude that, on one hand, $\Bd^\g$ is mathematically different from the operators previously considered in the literature and, on the other hand, we can restrict the domain of $\Bd^\g$ from $\cD'$ to $\zeta'$ and we can work in the space of ultra-distributions.


\subsection{Unitarity}\label{ultrun}

The contour prescription in Fig.~\ref{fig3} was systematically used to study the fractional d'Alem\-bertian in QFT restricting its domain to the space of ultra-distributions $\zeta'$ \cite{Barci:1996br,Barci:1998wp}. This contour produces the principal-value propagator \Eq{pvG} given by the average of the causal and anti-causal prescription. The authors of \cite{Bollini:1998hj,Plastino:2017kgl} studied the diagrammar obtained with the usual Feynman rules but replacing the Feynman propagator with the principal-value propagator \Eq{pvG}, which they called Wheeler propagator. While this procedure works at the tree level to eliminate the particle associated with $\tilde G_{\rm PV}$ from the physical spectrum, at higher orders in the loop expansion the optical theorem is violated and both renormalizability and unitarity bump into problems \cite{Anselmi:2020tqo}. The above procedure is an interesting precursor of the Anselmi--Piva prescription, which is more than a simple replacement of the propagator.

These considerations allow us to conclude the comparison between the approach with ultra-distributions and ours. In the former \cite{Barci:1996br,Barci:1998wp}, the domain of the fractional d'Alem\-bertian is restricted in a way naturally leading to an empty spectrum at the tree level \cite{Barci:1996ny} but also to a problematic QFT at higher loops \cite{Anselmi:2020tqo}. In our approach, we defined a different fractional d'Alembertian and let its domain span the space $\cD'$ of generalized functions but we can also restrict it to $\zeta'$. Since our propagator is not the principal-value one $\tilde G_{\rm PV}$, to remove ghosts we apply the Anselmi--Piva prescription, thus guaranteeing unitarity at all orders in perturbation theory. On the other hand, the ultra-distributions approach might be more convenient when considering other types of fractional d'Alembertians in classical systems, especially when one wants to perform a Hamiltonian analysis.


\section{Initial conditions in fractional dynamics}\label{sec6}

The problem of initial conditions in nonlocal dynamics cannot be treated in the usual way due to the presence of infinitely many derivatives. While several methods exist to make sense of linear nonlocal equations of motion \cite{Pais:1950za,Barnaby:2007ve}, the presence of non-linear interactions makes the problem much more difficult \cite{Eliezer:1989cr,Moeller:2002vx}. The diffusion method is a way to find consistent solutions (exact or approximate) of non-linear nonlocal equations of motion for certain nonlocal operators. Originally proposed to handle the tachyon dynamics in the low-energy limit of string field theory \cite{Forini:2005bs,Calcagni:2007wy,Calcagni:2007ru,Calcagni:2007ef,Mulryne:2008iq,Calcagni:2009jb}, the diffusion method has been adapted to the form factors in nonlocal QFTs and NLQG \cite{Calcagni:2018lyd,Calcagni:2018gke,Calcagni:2018fid}. Here we extend its application to fractional QFTs and FQG. Since we use the representation \Eq{Bdot}, the closest analogue in nonlocal QFT is the exponential form factor $\exp\B$, which is treated in \cite{Calcagni:2018lyd}. The reader can consult this reference for more details on the method.

The goal of the diffusion method is to trade the nonlocal operator in the dynamics with a translation in a fictitious extra coordinate, but in such a way as to have local $(D+1)$-dimensional equations of motion. The latter can be obtained only through a diffusion equation where the diffusion derivative operator is local (e.g., $\B$ or $\B^n$) and such equation can be obtained only if one can represent the original nonlocal operator in terms of another nonlocal operator with which we know how to diffuse. More technically, the diffusion method follows this logic (we assume a covariant model):
\begin{enumerate}
	\item Given a $D$-dimensional action $S$ with one or more nonlocal form factors $\cF(\B)$, one looks for a convenient parametric representation of $\cF(\B)$ in terms of one or more local (i.e., of finite derivative order) positive polynomials $A(\B)$. This is not possible in general but there are large classes of form factors which can be recast via the Balakrishnan--Komatsu representation as
	\ba
	\cF(\B)&=&a_0+\sum_{n=1}^N \left\{a_n^+ [A(\B)]^n+a_n^- [A(\B)]^{-n}\right\}\nn
	&=&a_0+\sum_{n=1}^N \left[a_n^+ A^n+\frac{a_n^-}{\G(n)}\int_0^{+\infty}\rmd \t\, \t^{n-1}\rme^{-\t A}\right],\label{F}
	\ea
	where $N$ can also be infinity. The form factors $\cF(\B)$ appearing in FQG and other quantum field theories \cite{Calcagni:2018lyd} are of this type, although in some cases it is necessary to manipulate $\cF$ more vigorously to obtain one or more polynomials $A(\B)$ \cite{Calcagni:2018gke}.
	\item Second, one rewrites the nonlocal action $S$ as an artificial $(D+1)$-dimensional system $\cS$ such that its fields $\Phi(r,x)$ diffuse according to one or more source-less diffusion equations $[\p_r+A(\B)]\Phi=0$, where $r$ is the extra direction. By construction, the equations of motion of the $(D+1)$-dimensional fields have the same form of the equations of motion from $S$, where the action of $\cF(\B)$ on a field is recast as a translation along the extra direction $r$, $\cF(\B)\Phi(r,x)=\Phi(r+\tilde r,x)$; $\tilde r$ could be integrated depending on the form of \Eq{F}. Due to this, the $(D+1)$-dimensional system is nonlocal in $r$ since fields are evaluated at different points in $r$, but is 
	local in spacetime derivatives. For this reason, we call $\cS$ \emph{localized}.
	\item Third, one solves the diffusion equation(s) with initial condition at $r=0$ such that $\Phi(0,x)$ solves the equations of motion of the \emph{local} system coming from $S$ when the
	nonlocality is switched off. Thus, one needs to specify only a finite number of initial conditions $\Phi(0,t_{\rm i},\bm{x})$, $\dot\Phi(0,t_{\rm i},\bm{x})$, $\dots$, to solve the diffusion equation.
	\item Once $\Phi(r,x)$ is found, one plugs it into the equations of motion and finds a value $r_*$ such that these equations are satisfied for all $x$. It is not guaranteed that such value exists but, if it does, then $\Phi(r_*,x)$ is a solution of the original nonlocal system $S$.
\end{enumerate}
Below, we carry out items 1-3 explicitly, leaving number 4 for the future.


\subsection{Diffusion method}

We take the example of a massless scalar field on Minkowski spacetime with potential $V(\phi)$ and action
\be\label{acts}
S=\int\rmd^Dx\left\{\frac12\,\phi\left[\B+\lst^{-2}(\lst^2\Bd)^\g\right]\phi-V(\phi)\right\},
\ee
where $\g>1$ and $\Bd^\g$ is given by the representation \Eq{Bdot}, which we recast here for a dimensionless parameter $\t$:
\ba
\hspace{-.7cm} (\lst^2\Bd)^\g &=&(\lst^4\B^2)^\frac{\g}{2}\nn
&=& \dfrac{1}{\Gamma(n-\g/2)}\displaystyle\int_0^{+\infty}\rmd\tau\,\tau^{n-\frac{\g}{2}-1}\,(\lst^4\B^2)^n\rme^{-\t\lst^4\B^2},\qquad n-1<\frac{\g}{2}<n\,.\label{Bdot2}
\ea
Since $\g>1$ for the $\Bd^\g$ term to dominate in the ultraviolet, we have $n\geq 1$. The equation of motion from \Eq{acts} is
\be\label{pheom}
\left[\B+\lst^{-2}(\lst^2\Bd)^\g\right]\phi-V'(\phi)=0\,.
\ee

The representation \Eq{Bdot} is ideal for the diffusion method because it features the exponential operator $\exp(-\t\B^2)$, for which we do have a localized representation in terms of a local diffusion equation. We show step by step that the $(D+1)$-dimensional localized system associated with \Eq{acts} is
\bs\label{llz}
\ba
\hspace{-1cm}\cS&=&\int \rmd^D x\,\rmd r \left(\cL_\Phi+\cL_\chi+\cL_\la\right)\,,\label{act}\\
\hspace{-1cm}\cL_{\Phi}&=&\frac12\Phi(r,x)\B\Phi(r,x)+\frac12\lst^{-2}f(r)\,\Phi(r,x)(\lst^2\B)^{2n}\tilde\Phi(r,x)-V[\Phi(r,x)]\,,\label{locPh2}\\
\hspace{-1cm}\cL_{\chi}&=&-\frac12\frac{\lst^{-2}}{\Gamma(n-\g/2)}\int_0^{+\infty}\rmd\tau\,\tau^{n-\frac{\g}{2}-1} f(r-\tau)\!\int_0^{\tau} \rmd q\,\chi(r-q,x)(\p_{r'}+\lst^4\B^2)\Phi(r',x)\,,\nn\label{locch2}\\
\cL_\la &=&  \la(r,x)\left[\chi(r,x)-(\lst^2\B)^{2n}\Phi(r,x)\right],
\ea\es
where $r$ is an artificial extra ``direction'' (actually, with dimensionality $[r]=0$), $r'=r+q-\tau$, $\Phi(r,x)$, $\chi(r,x)$ and $\la(r,x)$ are $(D+1)$-dimensional scalar fields with dimensionality $[\Phi]=[\chi]=(D-2)/2$ and $[\la]=(D+2)/2$ and
\be\label{tildephi}
\tilde\Phi(r,x)\coloneqq\frac{1}{\Gamma(n-\g/2)}\int_0^{+\infty}\rmd\tau\,\tau^{n-\frac{\g}{2}-1}\Phi(r+\tau,x)\,,\qquad 2(n-1)<\g<2n\,,
\ee
with $[\tilde\Phi]=[\Phi]$. Note that, in $\cL_\chi$, the fields $\chi$ and $\Phi$ are translated in the $r$ direction by different amounts. For the time being, the $(D+1)$-dimensional scalar $\Phi(r,x)$ is unrelated to the $D$-dimensional field $\phi(x)$; their connection will be made explicit in \Eqq{rstcon1}. Also, the function $f(r)$ is introduced for reasons that will become clear later.

The system \Eq{llz} is nonlocal in $r$ but local in the spacetime coordinates $x$; more precisely, it is a higher-order derivative model of order $4n$. This might induce one to believe that the system is plagued by ghosts. However, in general ghosts in the $(D+1)$-dimensional system disappear when projecting the dynamics onto the $r=r_*$ slice \cite{Calcagni:2018lyd}. Moreover, we have seen above that the Anselmi--Piva prescription removes them order by order in perturbation theory, so that the unitarity problem is not of our concern in this section. 


The equation of motion $\de \cS/\de\chi=0$ are derived as follows. Ignore the $x$-dependence everywhere and define the functional variation for any field $f$ as $\de f(r)/\de f(\bar r)=\de(r-\bar r)$. We have
\ba
0 &=& \frac{\de\cS}{\de\chi(\bar r,\bar x)}\nn
&=&\la(\bar r)-\frac12 \int \rmd r\,\frac{\lst^{-2}}{\Gamma(n-\g/2)}\int_0^{+\infty}\rmd\tau\,\tau^{n-\frac{\g}{2}-1} f(r-\tau)\nn
&&\qquad\qquad\qquad\qquad\times\int_0^{\tau}\rmd q\,\de(r-q-\bar r)(\p_{r'}+\lst^4\B^2)\Phi(r')\nn
&=& \la(\bar r)-\frac12\frac{\lst^{-2}}{\Gamma(n-\g/2)}\int_0^{+\infty}\rmd\tau\,\tau^{n-\frac{\g}{2}-1}\nn
&&\qquad\qquad\qquad\qquad\times\int_{\bar r}^{\tau+\bar r} \rmd r\, f(r-\tau)(\p_{r'}+\lst^4\B^2)\Phi(r')\Big|_{r'=2r-\bar r-\tau},\label{inter2}
\ea
where we used the fact that the integration of the Dirac distribution in $q$ implies $0<q=r-\bar r<\tau$, hence $\bar r<r< \tau+\bar r$. Reparametrizing with $\rho= r-\bar r$, one gets an integral of the form $\int_0^{\tau} \rmd \rho\, f(\rho+\bar r-\t)\,F(2\rho+\bar r-r_*)$. If
\be\label{la0}
\la(r,x)=0\,,
\ee
then the integrand must be identically zero on shell for any integration range, since $\bar r$ is arbitrary and $\tau$ is integrated in its turn. This leads to the quartic-order diffusion equation
\be\label{difeq}
(\p_r+\lst^4\B^2)\Phi(r,x)=0\,.
\ee
The condition \Eq{difeq} means that we can recast the action of the operator $\exp(-\tau\lst^4\B^2)$ as a translation in the $r$ coordinate, 
\be\label{trans}
\rme^{-\tau\lst^4\B^2}\Phi(r,x)=\rme^{\tau\p_r}\Phi(r,x)=\Phi(r+\tau,x)\,.
\ee
Therefore, from \Eqq{tildephi} one has 
\be\label{eomdot}
(\lst^2\B)^{2n}\tilde\Phi(r,x)=(\lst^2\Bd)^\g\Phi(r,x)\,.
\ee
A non-trivial $\la\neq 0$ would give rise to a source term in the diffusion equation \Eq{difeq} and would spoil the translation rule \Eq{trans}; hence the necessity of enforcing \Eq{la0} by hand. This highlights the fact that not all solutions of the localized system \Eq{llz} are also solutions of the nonlocal system \Eq{acts}. The diffusion method does not work according to a holographic principle and there is no one-to-one correspondence between the dynamics in the bulk and the nonlocal dynamics at the boundary \cite{Calcagni:2018lyd}. This is simply a feature of the method, not a liability, since the $(D+1)$-dimensional system is nothing more than a mathematical contraption to solve the nonlocal dynamics.

The second equation of motion $\de\cS/\de\la=0$ is
\be\label{condix}
\chi(r,x)=(\lst^2\B)^{2n}\Phi(r,x)\,,
\ee
implying that $\chi$ obeys the same diffusion equation as $\Phi$.

The third equation of motion $\de \cS/\de\Phi=0$ can be found after noting that
\ba
\hspace{-1cm}\cL_{\chi}&=&\frac12\frac{\lst^{-2}}{\Gamma(n-\g/2)}\int_0^{+\infty}\rmd\tau\,\tau^{n-\frac{\g}{2}-1}f(r-\t)\nn
\hspace{-1cm}&&\qquad\times\int_0^{\tau}\rmd q\Big\{-\p_{q}[\chi(r-q)\Phi(r')]+\Phi(r')(\p_{r'}-\lst^4\B^2)\chi(r-q)\Big\}\nn
\hspace{-1cm}&=&\frac12\frac{\lst^{-2}}{\Gamma(n-\g/2)}\int_0^{+\infty}\rmd\tau\,\tau^{n-\frac{\g}{2}-1}f(r-\t)\nn
\hspace{-1cm}&&\qquad\times\left\{[\chi(r)\Phi(r-\tau)-\chi(r-\tau)\Phi(r)]+\int_0^{\tau} \rmd q\,\Phi(r')(\p_{r'}-\lst^4\B^2)\chi(r-q)\right\},\nn\label{inpa2}
\ea
where we imposed \Eq{la0}. Since
\ba
\int\rmd r\,\frac12 f(r)\,\lst^{2(2n-1)}\Phi(r)\B^{2n}\frac{\de\tilde\Phi(r)}{\de\Phi(\bar r)}&\!\!\!=\!\!\!&\int\rmd r\,\frac12 f(r)\,\frac{\lst^{2(2n-1)}}{\Gamma(n-\g/2)}\nn
&\!\!\!\!\!\!&\times\int_0^{+\infty}\rmd\tau\,\tau^{n-\frac{\g}{2}-1}[\B^{2n}\Phi(r)]\frac{\de\Phi(r+\tau)}{\de\Phi(\bar r)}+O(\N)\nn
&\!\!\!=\!\!\!&\frac12\frac{\lst^{2(2n-1)}}{\Gamma(n-\g/2)}\nn
&\!\!\!\!\!\!&\times\int_0^{+\infty}\rmd\tau\,\tau^{n-\frac{\g}{2}-1}f(\bar r-\t)\,\B^{2n}\Phi(\bar r-\tau)+O(\N)\,,\nn\label{medio}
\ea
where $O(\N)$ are total derivatives, we reach the expression
\ba
0=\frac{\de\cS[\Phi,\chi]}{\de\Phi(\bar r,\bar x)}&=& \B\Phi(\bar r)-V'[\Phi(\bar r)]-(\lst^2\B)^{2n}\la(\bar r)
+\frac12\frac{\lst^{-2}}{\Gamma(n-\g/2)}\int_0^{+\infty}\rmd\tau\,\tau^{n-\frac{\g}{2}-1}\nn
&&\times\left\{f(\bar r)\,[(\lst^2\B)^{2n}\Phi(\bar r+\tau)+\chi(\bar r+\tau)]\vphantom{\int_{\bar r}^{\bar r}}\right.\nn
&&\qquad+f(\bar r-\t)\,[(\lst^2\B)^{2n}\Phi(\bar r-\tau)-\chi(\bar r-\tau)]\nn
&&\qquad\left.-\int_{\bar r}^{\bar r+\tau} \rmd r\,(\p_{-\bar r}+\lst^4\B^2)\chi(2r-\bar r-\tau)\right\}\,.\label{inter22}
\ea
Enforcing \Eq{la0} and \Eq{condix}, \Eqq{inter22} becomes
\be\label{loceom}
\B\Phi(r,x)+f(r)\,\lst^{-2}(\lst^2\B)^{2n}\tilde\Phi(r,x)-V'[\Phi(r,x)]=0\,,
\ee
which can also be written as
\be\label{loceom2}
\left[\B+f(r)\,\lst^{-2}(\lst^2\Bd)^\g\right]\Phi(r,x)-V'[\Phi(r,x)]=0\,,
\ee
thanks to \Eq{eomdot}. Fixing a slice $r=r_*$ where
\be\label{rstcon1}
\Phi(r_*,x)=\phi(x)\,,\qquad f(r_*)=1\,,
\ee
\Eqq{loceom2} reduces to \Eq{pheom}. Therefore, on this $r$-slice the \emph{ad-hoc} system \Eq{llz} reproduces the nonlocal dynamics of the original model. Note that we have not yet specified the function $f(r)$ except for its value at $r=r_*$.

To recapitulate, we showed that the localized system \Eq{llz} with the conditions \Eq{la0} and \Eq{rstcon1} is equivalent to the fractional nonlocal system \Eq{acts} on a certain slice $r=r_*$. The diffusion method neither fixes the value $r_*$ nor guarantees that such a slice exists and one must find $r_*$ by hand. In the best possible scenario, an $r_*$ exists such that the solution $\Phi(r,x)$ of the diffusion equation is an exact solution of the localized dynamics \Eq{loceom} at $r=r_*$. In a less ideal situation, the solution $\Phi(r_*,x)$ is approximate to a good degree. In the worst case, no diffusing solution of \Eq{loceom} can be found.


\subsection{General diffusing solution}\label{gendifsol}

The solutions of the diffusing system can be found according to the same recipe detailed in \cite{Calcagni:2018lyd}. First of all, we solve the diffusion equation \Eq{difeq} with a Dirac delta as the initial condition at $r=0$,
\be\label{difG}
(\p_r+\lst^4\B^2_x)\cG(r\lst^4,x-x')=0\,,\qquad \cG(0,x-x')=\de^D(x-x')\,,
\ee
where we highlighted the fact that the solution depends on the dimensionful combination $\varrho=r\lst^4$. For example, in $D=4$ Minkowski spacetime the solution is a Meijer G-function (\ref{appD}):
\ba
\cG(r\lst^4,x-x')&=&\frac{(s-s')^2}{2^{10}\pi^2 (r\lst^4)^{\frac32}}\,G_{03}^{20}\left[\left.\frac{(s-s')^4}{256\,r\lst^4}~\right|~\begin{matrix}\\-1,0,-\frac12\end{matrix}\right],\nn
(s-s')^2&\coloneqq& -(t-t')^2+|\bm{x}-\bm{x}'|^2\,.\label{solG}
\ea
This allows us to find the solution for any initial condition $\Phi(0,x)$ as
\be\label{solgen}
\Phi(r,x)=\int\rmd^Dx'\,\cG(r\lst^4,x-x')\,\Phi(0,x')\eqqcolon \Phi_{\rm diff}(r\lst^4,x)\,.
\ee


\subsection{Degrees of freedom}

The $(D+1)$-dimensional localized real scalar field theory \Eq{llz} has two scalar degrees of freedom $\Phi$ and $\chi$. On the $r=r_*$ slice (if it exists) where the system is equivalent to the $D$-dimensional fractional scalar field theory \Eq{acts}, the degree of freedom $\chi$ is no longer independent and the fractional theory has one non-perturbative scalar degree of freedom $\phi$.


\subsection{Initial conditions}\label{inicon}

We are now in a position to count the number of degrees of freedom and the number of initial conditions in the Cauchy problem.

The Cauchy problem on spacetime slices of the $(D+1)$-dimensional localized theory \Eq{llz} is specified by $4n+4$ $r$-dependent initial conditions $\Phi^{(p_1)}(r,t_{\rm i},\bm{x})$ and $\chi^{(p_2)}(r,t_{\rm i},\bm{x})$, $p_1=0,\dots,4n-1$, $p_2=0,1,2,3$. The constraint \Eq{condix} reduces the total count by 4, down to $4n$ initial conditions. However, we can further drop the count down to two by an observation implicit in \cite{Calcagni:2018lyd,Calcagni:2018gke} but unnoticed therein.

We set the initial profile at the beginning of diffusion ($r=0$) to be
\be\label{filoc}
\Phi(0,x)=\phi_{\rm loc}(x)\,,
\ee
where $\phi_{\rm loc}$ is the solution of the local equation of motion obtained from \Eq{pheom} when $\lst=0$:
\be\label{eomloc}
\B\phi_{\rm loc}-V'(\phi_{\rm loc})=0\,.
\ee
Indeed, plugging \Eq{solgen} into \Eqq{loceom} and sending $\lst\to 0$ one finds ($x$-dependence omitted)
\ba
0&=&\left\{\B\Phi_{\rm diff}(r\lst^4)+f(r)\,\lst^{-2}(\lst^2\B)^{2n}\tilde\Phi_{\rm diff}(r\lst^4)-V'[\Phi_{\rm diff}(r\lst^4)]\right\}\Big|_{\lst=0}\nn
&=&\B\Phi(0)-V'[\Phi(0)]\,,\label{eomlst0}
\ea
which is \Eq{eomloc}. 

Once this profile is fixed, the general solution of the diffusion equation \Eq{difeq} is calculated via \Eq{solgen}. Finally, the function $\Phi(r)=\Phi_{\rm diff}(r\lst^4)$ is plugged into \Eq{loceom2} to get an expression that should vanish for all $x$ for some value $r=r_*$. As we said above, this last step is not guaranteed to work exactly in general. Depending on the choice of $f(r)$, $\Phi(0,x)$ is a solution of \Eq{loceom2} at $r=0$ while keeping $\lst$ fixed only if $f(0)=0$:
\ba
&&\left\{\B\Phi_{\rm diff}(r\lst^4)+ f(r)\,\lst^{-2}(\lst^2\B)^{2n}\tilde\Phi_{\rm diff}(r\lst^4)-V'[\Phi_{\rm diff}(r\lst^4)]\right\}\Big|_{r=0}\nn
&&\qquad\qquad\qquad\qquad= \B\Phi(0)+f(0)\,\lst^{-2}(\lst^2\B)^{2n}\tilde\Phi(0)-V'[\Phi(0)]\,.\label{eomr0}
\ea
The choice of an $f$ such that $f(0)=0$ and of \Eq{eomlst0} as the definitory equation for $\Phi(0)$ is practically mandatory because \Eq{eomr0} with $f(0)\neq 0$ is nonlocal and requires prior knowledge of $\Phi(r)$ to solve the dynamics for $\Phi(0)$, which is absurd \cite{Eliezer:1989cr,Moeller:2002vx}. Also, \Eq{eomlst0} is compatible with the expectation that, perturbatively in $\lst$ (small nonlocality), the nonlocal form factor $\Bd^\g$ deforms the local solutions in such a way as to preserve the asymptotic behaviour at $x^\mu\to\pm\infty$. The diffusion method, however, is non-perturbative and works for any value of $\lst$.

To summarize, $\lst$ is fixed throughout the diffusion process and the system flows from the $r=0$ slice where $\Phi_{\rm diff}(0)$ is a solution of \Eq{eomlst0} to the $r=r_*$ slice (if it exists) where $\Phi_{\rm diff}(r_*\lst^4)$ is a solution of \Eq{loceom}. It is easy to confuse the $\lst=0$ condition \Eq{eomlst0} with the $r=0$ condition \Eq{eomr0} because the diffusing solution $\Phi_{\rm diff}$ depends on the product $\varrho=r\lst^4$, so that $\Phi(0)$ is the same in both limits. However, the nonlocal term in the equation of motion \Eq{loceom2} does not depend on the quantity $\varrho$ and the degeneracy of the limits $\lst\to 0$ and $r\to 0$ is broken, unless we fix $f(r)$ precisely to make the nonlocal term proportional to a positive power of $\vr$. Given that this term is proportional to $\lst^{2(2n-1)}$, we set
\be
f(r)=\left(\frac{r}{r_*}\right)^{\frac{2n-1}{2}}h(r)\,,
\ee
where $h(r)$ is regular at $r=0$ and $h(r_*)=1$. For instance, $h(r)\equiv 1$. The same subtle point is present in QFTs with entire (in particular exponential and asymptotically polynomial) form factors \cite{Calcagni:2018lyd,Calcagni:2018gke}. Therefore, one should bear in mind that the limits $r\to 0$ and $\lst\to 0$ do not give the same result if $f(0)\neq 0$.

Equation \Eq{filoc} is the key to the problem of initial conditions. It states that the Cauchy problem of the localized system \Eq{llz} formally has the same infinite indeterminacy as the original nonlocal system \Eq{acts}, encoded in the arbitrary function $\phi_{\rm loc}(x)$. However, this function is not completely arbitrary but is, in fact, determined by the local equation of motion \Eq{eomloc}. Therefore, once we choose the two initial conditions of \Eq{eomloc}, we obtain $\Phi(0,x)$ through \Eq{filoc}; from $\Phi(0,x)$, we get the profile $\Phi(r,x)$ via \Eq{solgen}; plugging $\Phi(r,x)$ into \Eq{loceom} and Taylor expanding in $x$, we impose the $r$-dependent coefficients of the expansion to vanish. If this happens for a certain value $r=r_*$, then we obtain the solution $\phi(x)=\Phi(r_*,x)$ of the nonlocal system \Eq{acts}.

In conclusion, the non-perturbative Cauchy problem of the fractional real scalar field theory \Eq{acts} is specified by two initial conditions
\be\label{phidotphi}
\phi(t_{\rm i},\bm{x})\,,\qquad \dot\phi(t_{\rm i},\bm{x})\,,
\ee
independently of the value of $\g$. Consistently, the only solutions of a purely nonlocal model $\cL=\phi|\lst^2\B|^\g\phi-V(\phi)$ with no kinetic term in the local limit are those satisfying $V'(\phi)=0$ and there are no initial conditions.

Since the diffusion method for nonlocal tensor fields with higher spin follows exactly the same steps (including a generalization of \Eq{filoc} to more fields with higher rank; see \cite{Calcagni:2018lyd,Calcagni:2018gke} for different types of nonlocal form factors), from \Eq{F} it is not difficult to convince oneself that the Cauchy problem \Eq{phidotphi} is valid more generally for each field in any nonlocal theory with ultraviolet nonlocality where the associated local system ($\lst=0$) is second-order in time derivatives. In particular, in NLQG with asymptotically polynomial operators \cite{Buoninfante:2022ild,BasiBeneito:2022wux,Koshelev:2023elc,Calcagni:2018gke} and in
fractional gravity \cite{Calcagni:2021ipd,Calcagni:2022shb,Calcagni:2021aap} the local limit of the theory is Einstein gravity and there are $D(D+1)$ initial conditions 
\be\label{gdotg}
g_{\mu\nu}(t_{\rm i},\bm{x})\,,\qquad \dot g_{\mu\nu}(t_{\rm i},\bm{x})\,.
\ee
If the associated local system has $N$ derivatives, then the number of initial conditions required per tensor component is $N$.



\subsection{Other self-adjoint extensions}

The operator \Eq{Bdot} is not the only self-adjoint extension of the fractional d'Alem\-bertian; the Fresnel representation \Eq{cos} is indeed an alternative. They both stemmed from the Balakrishnan--Komatsu representation, which is a concrete instance (among others, such as the semi-group representation \cite{Boc49,Phi52} cited in the introduction or the spectral or Källén--Lehmann representation used in Section~\ref{sec4}) of the very general Dunford integral \Eq{dunf}. Although, under standard hypotheses such as positivity, different representations may coincide on their common domain and define the same closed operator (as in the case of the fractional Laplacian $(-\N^2)^\g$), if one weakens or modifies these hypotheses (as in the case of the fractional d’Alembertian) different constructions may fail, produce different closures or extensions, or raise domain issues.\footnote{A well-known example of inequivalent representations in nonlocal QFT is the Wataghin form factor $\exp\B$. Its space of solutions differs greatly depending on whether it is written as a Taylor series $\exp\B=\sum_{n=0}^\infty \B^n/n!$ or as a translation in an extra coordinate in the diffusion method \cite{Calcagni:2007ru}.} Even so, the physics can be similar or identical. As is typical in nonlocal QFTs, there seldom are uniqueness results about specific operators but, in contrast, classes of operators usually share the same physics. Fractional QFT seems to be no exception. For instance, one can consider an infinite family of self-adjoint extensions $(\B^{2n})^{\g/(2n)}$ of the fractional d’Alembertian within the same Balakrishnan--Komatsu representation by exponentiating an arbitrary even power of $\B$. It is not difficult to check that, for finite $n$, this class of operators has the same general physical properties as the $n=1$ case.

As another example of robustness of some key physical properties in the class of self-adjoint fractional d'Alembertians, we argue that the counting of initial conditions is the same if we employ the operator \Eq{cos}. At the level of the localized equations of motion, the main differences is in the diffusion equation \Eq{difeq} and in the definition of the dressed field \Eq{tildephi}. The former is replaced by the second-order ordinary diffusion equation
\be\label{difeqcos}
(\p_r+\lst^2\B)\Phi(r,x)=0\,,
\ee
which leads to an imaginary shift in the extra direction when acting with a phase-like operator: $\rme^{\pm\rmi\tau\lst^2\B}\Phi(r,x)=\Phi(r\pm\rmi\tau,x)$. This implies a different dressed field \Eq{tildephi} given by
\ba
\tilde\Phi(r,x)&=&\frac{1}{\Gamma(n-\g/2)}\int_0^{+\infty}\rmd\tau\,\tau^{n-\frac{\g}{2}-1}\frac12\left[\Phi(r+\rmi\tau,x)+\Phi(r-\rmi\tau,x)\right]\nn
&=&\frac{1}{\Gamma(n-\g/2)}\int_0^{+\infty}\rmd\tau\,\tau^{n-\frac{\g}{2}-1}\Re\,\Phi(r+\rmi\tau,x)\,.\label{tildephicos}
\ea
The action formulation \Eq{llz} must be adapted to this structure (for example, by modifying $\cL_\chi$ in order to eventually get \Eq{loceom}) but, once this is done consistently, all the discussion in Section~\ref{inicon} is clearly the same, including the number of initial conditions \Eq{phidotphi}. The general diffusing solution \Eq{solgen} is modified and, as said above, the alternative representation \Eq{cos} is expected to have a different space of solutions, which may affect some details of the physics. We will not explore this further here.


\section{Conclusions}\label{concl}

Self-adjointness of the kinetic term in a QFT is a fundamental requirement to have a Hermitian action. In general, the latter is a necessary (albeit not sufficient) condition to have a stable Minkowski vacuum (convergence of the path integral, which we already know how to build for a fractional QFT \cite{Calcagni:2024xku}) and a unitary evolution in a closed quantum system without any tailored implementation of PT-invariance (see Section~\ref{sec3b}). In this paper, we defined the operator ``$|\B|^\g$" in such a way as to satisfy the requirements of self-adjointness and convenience for the problem of initial conditions. After examining different representations, we singled out \Eq{Bdot} as the most suitable for the construction of a fractional QFT with viable physical properties. A model with a purely fractional kinetic term has zero initial conditions and an empty spectrum both classically and at the quantum level at any given perturbative order (there are no real poles and complex ghost modes are projected out of the space of asymptotic states).

The spectrum of the full theory is of course populated, since the kinetic term is not purely fractional and has an infrared component $\B$ that provides the physical degrees of freedom, which are the usual ones. Indeed, we have also established that the number of initial conditions of fractional field theories with kinetic terms of the form ``$\B+|\B|^\g$'' is two for each field contributing to the physical spectrum. This conclusion holds not only for the operator \Eq{Bdot} but also for \Eq{cos} and, even more generally, for any nonlocal field theory being second-order in time derivatives when nonlocality is switched off, thus amending the over-counting of initial conditions of \cite{Calcagni:2018lyd,Calcagni:2018gke}. The advantages of the diffusion method are multiple: it can be applied to any system with or without field interactions and to any rank-$n$ tensor field, including the graviton \cite{Calcagni:2018lyd,Calcagni:2018gke}. However, the method also suffers from limitations, the main ones being that it does not contain in itself the guarantee to find exact or approximate solutions (see item 4 at the beginning of Section \ref{sec6}) and that, when no viable solution is found, it does not allow one to conclude much about the existence of such solutions. Because of this and of the intrinsic difficulty of the subject of nonlocal theories, the employment of the diffusion method has so far been reduced to the formal counting of initial conditions in certain theories plus a handful of explicit examples of solutions in scalar models, string field theory and quantum gravity.

The results found here have their primary application in fractional QFT, a framework proposed to construct a consistent perturbative theory of quantum gravity based on dimensional flow, a property of spacetimes appearing across different theories in the landscape of quantum gravity \cite{Calcagni:2016azd,Carlip:2019onx} and such that geometry and correlation functions change with the probed scale. By updating fractional QFTs in general and FQG in particular to the self-adjoint formulation built here, we can place these on a more solid ground and make rapid advances in different directions. The counting of initial conditions done here for the $\B+|\B|^\g$ kinetic term applies immediately to FQG and opens the way to the construction of solutions of the classical equations of motion. The latter can be found via the Balakrishnan--Komatsu representation \Eq{Bdot} and would replace the expressions obtained for the non-Hermitian version of the theory \cite{Calcagni:2021aap}. Unitarity with the mixed $\B+|\B|^\g$ operator will require a dedicated study \cite{CaBr} that will upgrade the analysis done in \cite{Calcagni:2021ljs,Calcagni:2022shb} for QFTs with a $(-\B)^\g$ kinetic term. Interestingly, the operator \Eq{Bdot} can also serve as the basis for defining spacetime fractional differential forms as done in \cite{LaNave:2017nex} for spatial slices with the fractional Laplacian $(-\N^2)^\g$. This would allow us to return to the program of \cite{Calcagni:2011kn,Calcagni:2011sz} and construct a purely fractional integro-differential structure of spacetime, leading to a radically different field theory with respect to the one with standard Lebesgue measure presented here and in \cite{Calcagni:2021aap,Calcagni:2021ljs,Calcagni:2022shb}.

Another application of our findings is of interest for the Caffarelli--Silvestre extension theorem \cite{Caffarelli:2007eci}. The latter establishes an equivalence between a $D$-dimensional Euclidean scalar field theory with fractional Laplacian as a kinetic term and a $(D+1)$-dimensional theory with a canonical kinetic term and a non-trivial measure weight in the extra direction:
\be
\frac12\int\rmd^Dx\,\phi(x)\,(-\N^2)^\g\phi(x)=-\frac{2^{2(\g-1)}\G(\g)}{\g\G(-\g)}\int\rmd^Dx\int_0^{+\infty}\rmd y\,y^{1-2\g}\,\p_M\Phi(x,y)\,\p^M\Phi(x,y),\label{corres}
\ee
where $M=1,\dots,D,D+1$ \cite{Caffarelli:2007eci}. This extension from a nonlocal lower-dimensional theory to a local higher-dimensional one has been generalized to $p$-forms \cite{LaNave:2017nex} and applied to conformal invariance in the Ising model \cite{Paulos:2015jfa}, fractional electromagnetism \cite{LaNave:2019mwv}, quantization of fractional scalar QFTs \cite{Frassino:2019yip}, entanglement entropy \cite{Basa:2019ywr} and fractional gauge theories \cite{Phillips:2019qkc}. In some of these settings, Wick rotation is assumed at one point or another of calculations, either promoting $(-\N^2)^\g\to(-\B)^\g$ directly in \Eq{corres} or Euclideanizing a Lorentzian system in order to apply \Eq{corres} \cite{LaNave:2019mwv,Frassino:2019yip,Basa:2019ywr}. However, this analytic continuation is generally not allowed in nonlocal QFT, since momentum integrals are performed on non-trivial paths (not necessarily closed) that do not allow a rigid Wick rotation in the usual sense \cite{Pius:2016jsl,Briscese:2018oyx,Chin:2018puw,BasiBeneito:2022wux,Calcagni:2021ljs}. In the case of fractional operators, this failure of the standard analytic continuation is obvious from the discussion of Section~\ref{sec3b}. It would therefore be important, on one hand, to check whether a Minkowski version of the Caffarelli--Silvestre theorem holds with our operator \Eq{Bdot}, for which the transition from one signature to the other occurs without incidents; and, on the other hand, to revisit the constructions of \cite{LaNave:2019mwv,Frassino:2019yip,Basa:2019ywr} in the light of this.

The self-adjoint representation proposed here and the diffusion method to solve the problem of initial conditions in the presence of fractional d'Alembertians may be of interest not only in the quantum-gravity community in relation to FQG and to other approaches such as the Caffarelli--Silvestre one \cite[Section 5]{Calcagni:2021ljs}
but also in other contexts as shown by recent developments and applications in condensed matter and fluid dynamics (e.g., \cite{Els22} and references therein). Such applications could actually constitute an interesting test bench for fractional theories since they deal with physical phenomena more manageable than those described by quantum gravity, in a way similar to what pursued in the analogue-gravity paradigm \cite{Barcelo:2005fc}. In particular, one could explore the advantages and drawbacks of the coordinate-independent diffusion method in solving nonlocal fractional systems compared to other coordinate-dependent methods \cite{Els22}.


\medskip

\noindent
\emph{Acknowledgments.} 
 G.C.\ is supported by grant PID2023-149018NB-C41 funded by the Spanish Ministry of Science, Innovation and Universities MCIN/AEI/10.13039/ 501100011033. We thank D.~An\-sel\-mi and L.~Rachwa\l\ for various invaluable discussions.


\appendix


\section{\Eq{Gz} is not a representation of \texorpdfstring{$|\B|^{-\g}$}{|Box|-g}}\label{appA}

In this appendix, we attempt to find \Eq{Gk2} starting from the Cauchy representation \Eq{Gz}. For that, we need to carefully define the complex function $\tilde G(z)$. We can start from the kinetic operator $[(-\B)^2]^{\g/2}$ instead of $(\B^2)^{\g/2}$; the difference is just a choice of phase for the variable $z$. 

Is $\tilde G(z)=[(-z)^2]^{-\g/2}$, $(-z)^{-\g}$, $(z^2)^{-\g/2}$, $z^{-\g}$ or something else? Recall the typical example of the multi-valued function $\sqrt{z^2}$. Calling $\theta=\Arg\,z$ (hence $-\pi<\theta\leq\pi$), we have $\Arg\,z^2=2\,\Arg\,z=2\theta$. If we take the square root of this, we get one root corresponding to the value of the principal square-root function $\sqrt{z^2}=z$. If, instead, we do not pick the principal argument of $z^2$, then $\arg(z^2)=2\theta+2\pi k$, where $k=0,1$, and the operation $\sqrt{\cdot}$ yields two different functions $(\sqrt{z^2})_k=z\,\rme^{\rmi\pi k}$, $(\sqrt{z^2})_0=z$ and $(\sqrt{z^2})_1=-z$. Similarly, complex powers are defined through the logarithm as $z^\a=\exp(\a\ln\,z)$ (where $\a\in\mathbb{C}$) and, if one takes the same branch $\ln\,z={\rm Ln}\,z+2\pi k\rmi$ consistently, one obtains the principal value of the power obeying the composition rule
\be
(z^\a)^\b = z^{\a\b}\,,\qquad z\neq 0\,.
\ee
Thus, provided $\b=n\in\mathbb{Z}$ or one consistently takes the same branch, the order of application of the powers $\a$ and $\b$ does not count.\footnote{For the $k$-th branch of the logarithm, one has $z^\a=\rme^{\a\ln\,z}=\rme^{\a({\rm Ln}\,z+2\pi k\rmi)}=\rme^{\a(\ln|z|+\rmi\theta+2\pi k\rmi)}=|z|^\a\rme^{\rmi\a\theta}\rme^{2\pi\a k\rmi}$. Then, $(z^\a)^\b=(|z|^\a\rme^{\rmi\a\theta}\rme^{2\pi\a k\rmi})^\b=|z|^{\a\b}\rme^{\rmi\a\b\theta}\rme^{2\pi\a\b k\rmi}$. On the other hand, $z^{\a\b}=|z|^{\a\b}\rme^{\rmi\a\b\theta}\rme^{2\pi\a\b k'\rmi}=(z^\a)^\b\rme^{2\pi\a\b (k'-k)\rmi}$ and the two are the same if one takes the same branch $k=k'$. Using the property $(\rme^z)^n=\rme^{zn}$ if $n\in\mathbb{Z}$, one can also show that $(z^\a)^n = z^{\a n}$.}

In the case of a physical QFT such as the one we are considering, the choice of branch is part of the definition of the theory and cannot be changed \emph{in media res}. In particular, one can take the principal branch of the logarithm ($k=0$), ${\rm Ln}\,z=\ln|z|+\rmi\theta$, $-\pi<\theta\leq\pi$. Then, the principal value of $-z$ is $0<\Arg(-z)=\theta+\pi\leq 2\pi$ and one must perform all the calculations in the Riemann sheet $S_0$ with phase $\in(0,2\pi]$, where
\be\label{Gbr0}
\tilde G(z)=(-z)^{-\g}\,,\qquad \textrm{branch } \sqrt{z^2}=z\,,
\ee
exactly the same as in \Eq{Gz1}. Then, the contour for the causal propagator is the one in Fig.~\ref{fig1} and the K\"allén--Lehmann representation is \Eq{propold}. If one takes the other branch of $\sqrt{z^2}$, then 
\be\label{Gbr1}
\tilde G(z)=z^{-\g}\,,\qquad \textrm{branch } \sqrt{z^2}=-z\,,
\ee
and calculations must be performed on the Riemann sheet spanned by a phase $\in(-\pi,\pi]$. The final result is the same and is wrong. The propagator \Eq{propold} is not even under the reflection $k^2\to-k^2$ as was the original expression \Eq{Gk2} and the associated theory has a non-Hermitian action. Therefore, the Cauchy integral \Eq{Gz} does not represent the propagator of the self-adjoint operator $(\B^2)^{\g/2}$ and, in particular, of its representation $\Bd^\g$ \Eq{Bdot}.


\section{Derivation of \Eq{KLfin} in the complex \texorpdfstring{$z$}{z}-plane}\label{appB}

The arcs at infinity in Fig.~\ref{fig2} give a zero contribution for $\Re\,\g>0$, as one can check with the parametrization $z=R\,\rme^{\rmi\theta}$ in the limit $R\to\infty$ \cite{Calcagni:2021ljs,Calcagni:2022shb}. The contributions $L_\pm$ along the imaginary axis are parametrized by some $t$ running from $\pm\rmi\infty$ to $\mp\rmi\infty$:
\ba
\tilde G_\pm(-k^2) &=& \lim_{\ve\to 0^+}\frac{1}{2\pi\rmi}\int_{\pm\rmi\infty}^{\mp\rmi\infty}\rmd t\, \frac{\tilde G(t\pm\ve)}{t+k^2}\nn
&\stackrel{t=-\rmi s}{=}& \lim_{\ve\to 0^+}\frac{-1}{2\pi}\int_{\mp\infty}^{\pm\infty}\rmd s\, \frac{\tilde G(\pm\ve-\rmi s)}{k^2-\rmi s}\nn
&=& \lim_{\ve\to 0^+}\frac{\mp 1}{2\pi}\int_0^{+\infty}\rmd s \left[\frac{\tilde G(\pm\ve+\rmi s)}{k^2+\rmi s}+\frac{\tilde G(\pm\ve-\rmi s)}{k^2-\rmi s}\right],\label{Gpm}
\ea
where $\ve$ is sent to zero at the end of the calculation. Calling $\ve^\pm=\pm\ve$ and keeping track of two independent choices of sign at the same time, we have
\ba
\tilde G(\ve^\pm \pm\rmi s) &=& \lim_{\ve^\pm\to 0^\pm}[(\ve^\pm \pm\rmi s)^2]^{-\frac{\g}{2}}\nn
&=& \lim_{\ve^\pm\to 0^\pm} s^{-\g}\left(-1\pm\rmi\frac{2\ve^\pm}{s}\right)^{-\frac{\g}{2}}\nn
&=& s^{-\g}\rme^{-\rmi\frac{\g}{2}\vp_\pm^\pm},
\ea
where
\be
\vp_\pm^\pm\coloneqq \lim_{\ve^\pm\to 0^\pm} \Arg\left(-1\pm\rmi\frac{2\ve^\pm}{s}\right)= \lim_{\ve^\pm\to 0^\pm} \arctan\left(\frac{\pm2\ve^\pm/s}{-1}\right)+\de\,.\label{argde}
\ee
Here we used the following piecewise definition the principal value of the argument $\Arg$:
\be\label{argxy}
\Arg(x + \rmi y) =
\left\{\begin{matrix}
	\arctan\left(\frac{y}{x}\right)\hphantom{+\pi}\qquad \text{if }\, x > 0\,,\hphantom{\, y \geq 0\,,} \\
	\arctan\left(\frac{y}{x}\right) + \pi\qquad \text{if }\, x < 0 \,,\, y \geq 0\,, \\
	\arctan\left(\frac{y}{x}\right) - \pi\qquad \text{if }\, x < 0 \,,\, y < 0\,.
\end{matrix}\right.
\ee
In our case, the phase $\de$ in \Eq{argde} is $\de=\pi$ if $\pm\ve^\pm>0$ and $\de=-\pi$ if $\pm\ve^\pm<0$, so that
\be
\vp^+_+=\vp^-_-=\pi\,,\qquad \vp^+_-=\vp^-_+=-\pi\,.
\ee
Therefore, \Eqq{Gpm} becomes
\ba
\tilde G_\pm(-k^2) &=&\mp\frac{1}{2\pi}\int_0^{+\infty}\frac{\rmd s}{s^\g} \left[\frac{\rme^{-\rmi\frac{\g}{2}\vp_+^\pm}}{k^2+\rmi s}+\frac{\rme^{-\rmi\frac{\g}{2}\vp_-^\pm}}{k^2-\rmi s}\right]\nn
&=& \mp\frac{1}{2\pi}\int_0^{+\infty}\frac{\rmd s}{s^\g} \left[\frac{\rme^{\mp\rmi\frac{\pi\g}{2}}}{k^2+\rmi s}+\frac{\rme^{\pm\rmi\frac{\pi\g}{2}}}{k^2-\rmi s}\right]\nn
&=& \frac{1}{\pi}\int_0^{+\infty}\frac{\rmd s}{s^\g}\,\frac{s\,\sin\frac{\pi\g}{2}\mp k^2\cos\frac{\pi\g}{2}}{s^2+k^4}\,,\label{Gpm2}
\ea
and the total \Eq{Gtot} is
\be
\tilde G(-k^2)=\frac{1}{\pi}\int_0^{+\infty}\frac{\rmd s}{s^\g}\,\frac{s\,\sin\frac{\pi\g}{2}+|k^2|\cos\frac{\pi\g}{2}}{s^2+k^4}\,,
\ee
where we used $\Theta(-k^2)+\Theta(k^2)=1$ and $-\Theta(-k^2)+\Theta(k^2)={\rm sgn}(k^2)$. Noting that 
\ben
\tan\left(\frac{\pi\g}{2}\right)\int_0^{+\infty}\frac{\rmd s}{s^\g}\,\frac{s}{s^2+k^4}=\int_0^{+\infty}\frac{\rmd s}{s^\g}\,\frac{|k^2|}{s^2+k^4}\,,
\een
after a reparametrization $s'=s^2$ we obtain \Eq{KLfin}.

One can double check \Eq{Gpm2} by choosing the branch $\sqrt{z^2}=z$ for the propagator in the $\Re\,z>0$ half-plane and the branch $\sqrt{z^2}=-z$ for the propagator in the $\Re\,z<0$ half-plane. The gluing together of the two branches gives \Eq{Gtot}. According to \ref{appA}, in these branches $\tilde G(z)=(\sqrt{z^2})^{-\g}$ is split into two different functions:
\be
\tilde G(z) =\left\{\begin{matrix}
	\tilde G_+(z) =z^{-\g}\hphantom{(-)}\,,\qquad \Re\,z>0\\
	\tilde G_-(z) =(-z)^{-\g}\,,\qquad \Re\,z<0\\
\end{matrix}\right.\,.\label{Gz3}
\ee
Repeating the calculation of $\tilde G_\pm(-k^2)$ with \Eq{Gz3}, \Eqq{Gpm} is replaced by
\be
\tilde G_\pm(-k^2) = \lim_{\ve\to 0^+}\frac{\mp 1}{2\pi}\int_0^{+\infty}\rmd s \left[\frac{\tilde G_\pm(\pm\ve+\rmi s)}{k^2+\rmi s}+\frac{\tilde G_\pm(\pm\ve-\rmi s)}{k^2-\rmi s}\right].\label{Gpm3}
\ee
Then,
\ba
&&\tilde G_+(\ve\pm\rmi s)=(\ve\pm\rmi s)^{-\g}=s^{-\g}\,\rme^{-\rmi\g\,\arctan\frac{\pm s}{\ve}}\to s^{-\g}\,\rme^{\mp\rmi\frac{\pi\g}{2}}\,,\\
&&\tilde G_-(-\ve\pm\rmi s)=(\ve\mp\rmi s)^{-\g}=s^{-\g}\,\rme^{-\rmi\g\,\arctan\frac{\mp s}{\ve}}\to s^{-\g}\,\rme^{\pm\rmi\frac{\pi\g}{2}}\,,
\ea
so that \Eq{Gpm3} is indeed \Eq{Gpm2} and one reobtains \Eq{KLfin}.


\section{Transforming the complex plane}\label{appC}

In this rather technical appendix, we tackle the problem of finding the contour $\G^0\subset\{(\Re\,k^0,\,\Im\,k^0)\}$ given a contour $\G\subset\{(\Re\,z,\,\Im\,z)\}$ such that the first line of the identity
\ba
G(x-x')&=&\int_{\G^0\times \mathbb{R}^{D-1}} \frac{\rmd^Dk}{(2\pi)^D}\,\rme^{\rmi k\cdot (x-x')}\,\tilde G(-k^2)\nn
&=&\int_{\G^0\times \mathbb{R}^{D-1}} \frac{\rmd^Dk}{(2\pi)^D}\,\rme^{\rmi k\cdot (x-x')}\left[\frac{1}{2\pi\rmi}\int_{\G}\rmd z\,\frac{\tilde G(z)}{z+k^2-\rmi\e}\right]\label{equal}
\ea
is well-defined, where the prescription for $k^2$ is included to make the discussion general ($\e=0$ in the fractional case). In this formula, the unknown element to be determined is $\G^0$, which we want to compare with the path in Fig.~\ref{fig3}. The final $\G^0$ should have two characteristics. First, the $k^0$-integral must converg, which is dictated by Jordan's lemma: defining the Fourier anti-transform with an $\exp[\rmi k\cdot (x-x')]$ phase, $\G^0$ should be closed in the $\Im\,k^0<0$ half plane if $x^0-x^{\prime 0}>0$ and in the $\Im\,k^0>0$ half plane if $x^0-x^{\prime 0}<0$. Second, $\G^0$ should adhere to the real axis as much as possible, since $k^0\in\mathbb{R}$ in a physical situation. Jordan's lemma and maximal adherence of $\G^0$ to the $\Im\,k^0=0$ axis are sufficient to determine $\G^0$ uniquely, up to homotopic deformations.

The goal is to transform 
\be\label{Gzi}
\tilde G(-k^2)=\frac{1}{2\pi\rmi}\int_{\G}\rmd z\,\frac{\tilde G(z)}{z+k^2-\rmi\e}
\ee
to a representation on the complex $k^0$-plane and find the transformed contour $\G^0$. First, with a well-known manipulation we rewrite the Feynman prescription as 
\be
k^2-\rmi\e=-k_0^2+\om^2-\rmi\e\to -k_0^2+\om^2-2\rmi\om\e\simeq -k_0^2+(\om-\rmi\e)^2\eqqcolon-k_0^2+\tilde\om^2\,,
\ee
where $\om=\sqrt{|\bm{k}|^2+m^2}$ (with $m=0$ in this example). These are the poles $k^0=\pm\tilde\om=\pm\om\mp\rmi\e$ of, respectively, positive and negative energy below and above the real axis of the complex $k^0$-plane. Call 
\be
z^{\prime2}\coloneqq z+\tilde\om^2
\ee
and apply this change of variables to \Eqq{Gzi}:
\ba
\tilde G(k^0,\tilde\om) &=& \frac{1}{2\pi\rmi}\int_{\G'}\rmd z'\,\frac{2z'\tilde G(z^{\prime2}-\tilde\om^2)}{(z'-k^0)(z'+k^0)}\,.
\ea
Since the square root is multi-valued, we have to select a branch 
\be\label{z2bra}
z'_\pm\coloneqq\pm\sqrt{z+\tilde\om^2}=|z+\om^2|^\frac12\rme^{\frac{\rmi}{2}\,\Arg(z+\tilde\om^2)+\rmi m\pi}\,,\qquad m=0,1\,,
\ee
where $z'_\pm$ are two different functions and we sent $\e\to 0$ in the modulus. It turns out that, actually, we have to take both branches at the same time with a non-analytic operation. In fact, from \Eq{z2bra} we have that $\Im\,z'=0$ only when $\Arg(z+\tilde\om^2)=\arctan[(\Im\,z-2\rmi\e\om)/(\Re\,z+\om^2)]+\de=0$ (which is the only multiple of $2\pi$ in the interval $(-\pi,\pi]$), where the phase $\de=0,\pm\pi$ is given in \Eq{argxy}. Therefore, in the limit $\e\to 0$ we get $\Im\,z'=0$ if, and only if, $\Im\,z=0$, i.e., points of the real $z'$-axis can only be the image of points of the real $z$-axis. In that case, however, from \Eq{z2bra} we have that $z'=\pm\sqrt{|z+\om^2|}$ and each branch can only cover one semi-axis $z'\gtrless 0$. Therefore, we must take both branches \Eq{z2bra} simultaneously to cover the whole $z'$-plane, one for each half-plane:
\bs\label{transf}\ba
\Re\,z'>0\,:\qquad z'&=&z_+'=+\sqrt{z+\tilde\om^2}\,,\label{transf1}\\
\Re\,z'<0\,:\qquad z'&=&z_-'=-\sqrt{z+\tilde\om^2}\,.\label{transf2}
\ea\es
This is necessary to have \Eq{equal} cover both signs of $\Re\,k^0$, in such a way that
\be
\tilde G(k^0,\tilde\om)=\frac{1}{2\pi\rmi}\left[\Theta\left(\Re\,k^0\right)\int_{\G'_+}\rmd z'\,\frac{\tilde G_+'(z',\tilde\om^2)}{z'-k^0}+\Theta\left(-\Re\,k^0\right)\int_{\G'_-}\rmd z'\,\frac{\tilde G_-'(z',\tilde\om)}{z'+k^0}\right],\label{gpr}
\ee
where $z'=z_+'$, $\G_\pm'$ lie, respectively, in the $\Re\,z'\gtrless 0$ half-plane and
\be\label{Gprime}
G_\pm'(z',\tilde\om) = \frac{2z'}{\pm z'+k^0}\,\tilde G(z^{\prime2}-\tilde\om^2)\,.
\ee
We can thus rewrite \Eq{equal} as
\bs\label{equal2}\ba
G(x-x')&=& \int_{\G^0_+\times \mathbb{R}^{D-1}} \frac{\rmd^Dk}{(2\pi)^D}\,\rme^{\rmi k\cdot (x-x')}\tilde G'_+(k^0,\tilde\om^2)\nn
&&+\int_{\G^0_-\times \mathbb{R}^{D-1}} \frac{\rmd^Dk}{(2\pi)^D}\,\rme^{\rmi k\cdot (x-x')}\tilde G'_-(k^0,\tilde\om^2)\,,\\
\tilde G'_\pm(k^0,\tilde\om) &=&\frac{\Theta\left(\pm\Re\,k^0\right)}{2\pi\rmi}\int_{\G'_\pm}\rmd z'\,\frac{\tilde G_\pm'(z',\tilde\om)}{z'\mp k^0}\,.
\ea\es
If only one branch is picked, then by analytic continuation one can still cover most of the complex plane but with the exception of at least one real semi-axis (the branch cut of the $\sqrt{\cdot}$ function).

The recipe to find $\G_0$ is:
\begin{enumerate}
	\item[1.] Choose a convenient contour $\G$ in the $k^2$-plane to express $\tilde G(-k^2)$ as a Cauchy integral.
	\item[2.] Apply the transformations \Eq{transf} to obtain the contours $\G_+'$ and $\G_-'$.
	\item[3.] Calculate the residues in $\pm k^0$ to identify the above deformed contours with $\G^0$.
	\item[4.] Deform $\G^0$ so that it adheres to the real $k^0$-axis as much as possible.
\end{enumerate}
If needed, steps 3 and 4 can be interchanged and one deforms $\G_+'$ and $\G_-'$ before calculating the residue.


\subsection{Standard Feynman propagator}

The example of the standard causal propagator with Feynman prescription will help to understand how to make the comparison between results in the $k^2$-plane and in the $k^0$-plane. Consider the Cauchy representation of the standard massless propagator with Feynman prescription:
\be\label{fey}
\frac{1}{k^2-\rmi\e}=\frac{1}{2\pi\rmi}\int_{\G_{\rm F}}\rmd z\,\frac{\tilde G(z)}{z+k^2-\rmi\e}\,,\qquad \tilde G(z)=-\frac{1}{z}\,.
\ee

\begin{figure}[ht]
	\bc
	\includegraphics[width=12cm]{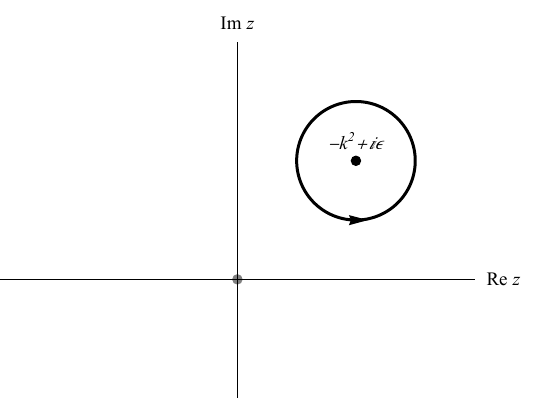}
	\ec
	\caption{\label{fig4} Non-maximal contour $\G_{\rm F}$ of the Cauchy representation \Eq{fey} of the standard Feynman propagator $1/(k^2-\rmi\e)$. The radius $r$ of the circle is arbitrary and small but does not have to be infinitesimal. The gray dot marks the pole of $\tilde G(z)$.}
\end{figure}

\emph{Step 1.} The contour $\G=\G_{\rm F}$ could be taken to be the maximal one, i.e., the one encompassing the maximum number of points in the complex case. This would be the same contour shown in Fig.~\ref{fig1}, where there is no branch cut on the positive real semi-axis. We applied the transformation \Eq{transf} to this maximal contour and checked that, indeed, we obtain the correct Feynman contour $\G^0_{\rm F}$ at the end. However, here we show a different route to the same result, using a most convenient non-maximal contour made of a counter-clockwise circle around $-k^2+\rmi\e$ of small (not necessarily vanishing) radius $r$ (Fig.~\ref{fig4}).

\begin{figure}[ht]
	\bc
	\includegraphics[width=8.5cm]{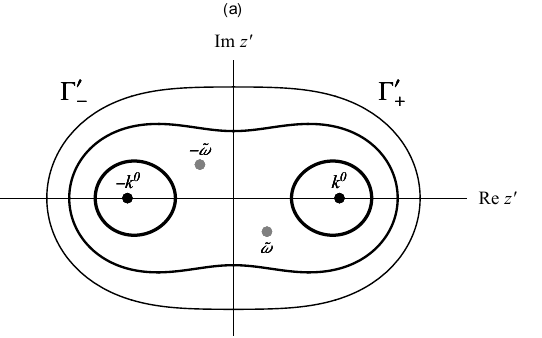}
	\includegraphics[width=8.5cm]{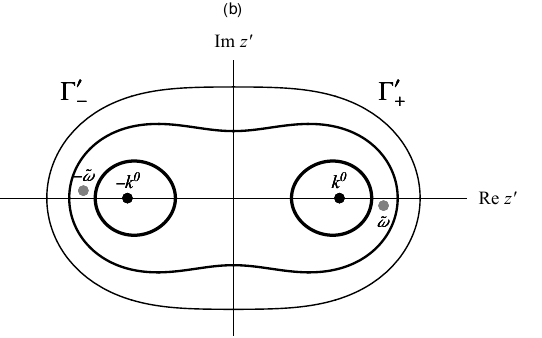}
	\ec
	\caption{\label{fig5} Transformed Feynman paths $\G'_+$ (curves in the $\Re\,z'>0$ half-plane) and $\G'_-$ (curves in the $\Re\,z'<0$ half-plane) for decreasing parameter $r$ (increasing thickness) in the cases (a) $|k^0|>\om$ (time-like $k^2<0$) and (b) $|k^0|<\om$ (space-like $k^2>0$), for the same value of $k^0$ and opposite value of $k^2$. Actually the paths do not cross the imaginary axis $\Re\,z'=0$ and points thereon are not part of $\G_+'\cup\G_-'$. All paths are counter-clockwise.}
\end{figure}

\emph{Step 2.} $\G_{\rm F}$ is parametrized by $z=r\,\rme^{\rmi\theta}$ with $r<|k^2|$. Applying \Eq{transf}, this circle is mapped onto one of the curves $z_\pm'=\pm (r\,\rme^{\rmi\theta}+k_0^2)^{1/2}$. When $r\gg |k^0|$, each curve is an open arc centered at the origin in the complex $z'$-plane. As $r$ is progressively reduced, each arc is dragged away from the origin until it closes on itself. When $r\ll |k^0|$, these curves become deformed loops centered at $\pm k^0$ and ovalized towards the origin (Fig.~\ref{fig5}). In their respective domain, the functions \Eq{Gprime}
\be
G_\pm'(z',\tilde\om)=\frac{2z'}{\pm z'+ k^0}\,\frac{1}{-z^{\prime2}+\tilde\om^2}
\ee
have two poles, one at $z'=\mp k^0$ and one at $z'=\pm\tilde\om$, also shown in Fig.~\ref{fig5} in the two cases $|k^0|>\om$ (time-like $k^2<0$) and $|k^0|<\om$ (space-like $k^2>0$). 

\begin{figure}[ht]
	\bc
	\includegraphics[width=12cm]{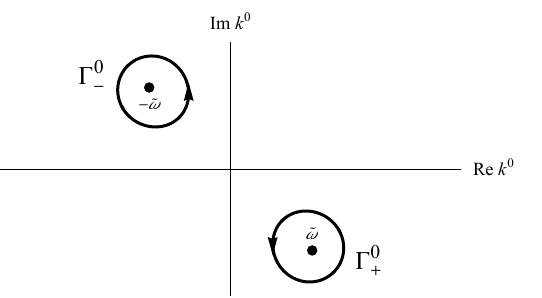}
	\ec
	\caption{\label{fig6} Paths $\G^0_+$ and $\G^0_-$ with the Feynman prescription in the $k^0$-plane.}
\end{figure}

\begin{figure}[ht]
	\bc
	\includegraphics[width=12cm]{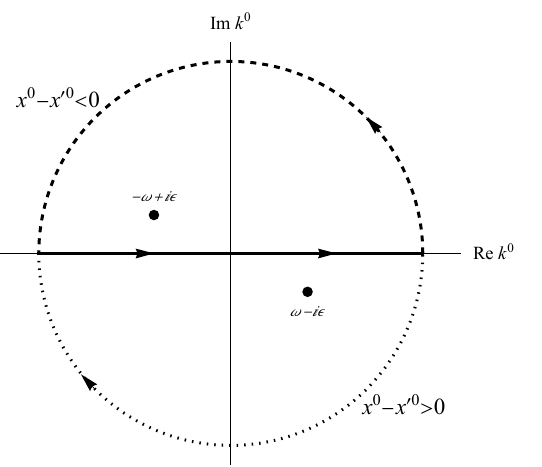}
	\ec
	\caption{\label{fig7} Standard Feynman contours $\G^0_{\rm F}$ in the $k^0$-plane. The arc at infinity is in the upper half-plane if $x^0-x^{\prime 0}>0$ (dashed semi-circle, from the deformation of $\G^0_-$) and in the lower half-plane if $x^0-x^{\prime 0}<0$ (dotted semi-circle, from the deformation of $\G^0_+$).}
\end{figure}

\emph{Step 3.} Let us take $r$ small enough so that $\G_+'$ and $\G_-'$ are disjoints and do not encircle $\pm\om$. Performing the Cauchy integrals in \Eq{equal2} on the transformed contours $\G_\pm'$ yields
\be
\tilde G'_\pm(k^0,\tilde\om) =\pm\frac{\Theta\left(\pm\Re\,k^0\right)}{-k_0^2+\tilde\om^2}=\pm\frac{\Theta\left(\pm\Re\,k^0\right)}{k^2-\rmi\e}\,,
\ee
so that
\ben
G(x-x')=\int_{\G^0_+\times \mathbb{R}^{D-1}}\frac{\rmd^Dk}{(2\pi)^D}\,\rme^{\rmi k\cdot (x-x')}\frac{\Theta\left(\Re\,k^0\right)}{k^2-\rmi\e} -\int_{\G^0_-\times \mathbb{R}^{D-1}}
\frac{\rmd^Dk}{(2\pi)^D}\,\rme^{\rmi k\cdot (x-x')}\frac{\Theta\left(-\Re\,k^0\right)}{k^2-\rmi\e}\,,
\een
where $\G^0_+$ and $\G^0_-$ are determined by the integration domains $\G_\pm'$ in the $z'=k^0$ direction: they are the thick ovals in Fig.~\ref{fig5} but in the $k^0$-plane, around $\pm\tilde\om$ and parametrized by $k^0=\pm(r\,\rme^{\rmi\theta}+\tilde\om^2)^{1/2}$ (Fig.~\ref{fig6}). From Fig.~\ref{fig6}, one has that $\Theta\left(\pm\Re\,k^0\right)$ corresponds to $\Theta\left(\mp\Im\,k^0\right)$ for small $r$, so that
\be
G(x-x')=\int_{\G^0_+\times \mathbb{R}^{D-1}}\frac{\rmd^Dk}{(2\pi)^D}\,\rme^{\rmi k\cdot (x-x')}\frac{\Theta\left(-\Im\,k^0\right)}{k^2-\rmi\e} -\int_{\G^0_-\times \mathbb{R}^{D-1}}
\frac{\rmd^Dk}{(2\pi)^D}\,\rme^{\rmi k\cdot (x-x')}\frac{\Theta\left(\Im\,k^0\right)}{k^2-\rmi\e}\,.\label{equal3}
\ee
\emph{Step 4.} At this point, we deform both contours so that they run along the whole real axis, plus an arc that is sent to infinity and gives a vanishing contribution thanks to Jordan's lemma ($\Theta\left(\mp\Im\,k^0\right)$ is equivalent to $\Theta\left[\pm(x^0-x^{\prime 0})\right]$). These are precisely the standard Feynman contours $\G^0_{\rm F}$ in ordinary QFT, depicted in Fig.~\ref{fig7}. 



\subsection{Purely fractional propagator}

This rather tortuous way to obtain the well-known result given in Fig.~\ref{fig7} serves as an example of how to find the path $\G^0$ from the contour in the Cauchy representation of a propagator. We now solve the same problem for the propagator of the self-adjoint operator $|\B|^\g$. 

\begin{figure}
	\bc
	\includegraphics[width=12cm]{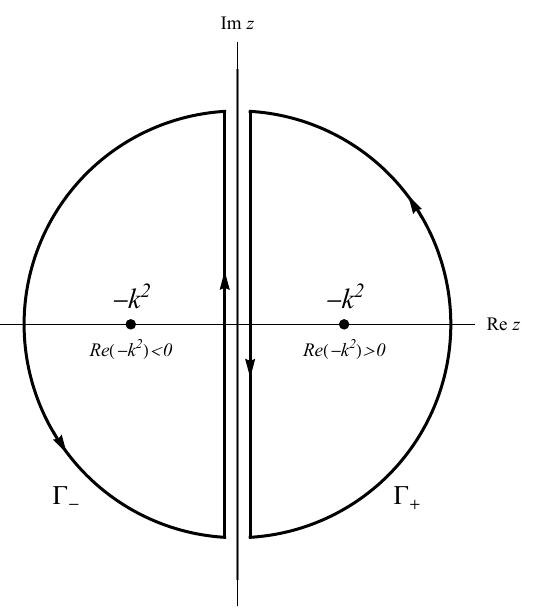}
	\ec
	\caption{\label{fig8} Maximal contour $\G=\G_+\cup\G_-$ of the Cauchy representation \Eq{Gtot} of the fractional propagator $(k^4)^{-\g/2}$. The vertical gray line marks the discontinuity of $\tilde G(z)$ on the imaginary axis.}
\end{figure}

\emph{Step 1.} Again, we can start from the maximal contour or from a non-maximal one. Contrary to the standard case, here it is more convenient to take the maximal contour (depicted in Fig.~\ref{fig2}), since it clearly defines the maximal domain of the two function $\tilde G_\pm(z)$. We checked that taking a non-maximal double contour leads to the same result but requires extra care and effort. Setting $\e=0$ in the pole, we repeat Fig.~\ref{fig2} in Fig.~\ref{fig8}.

\begin{figure}
	\bc
	\includegraphics[width=14cm]{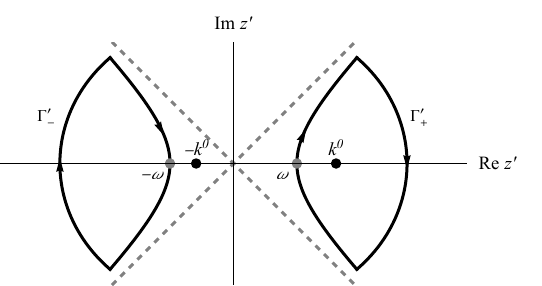}
	\ec
	\caption{\label{fig9} Transformed counter-clockwise contours $\G'_+$ ($\Re\,z'>0$ half-plane) and $\G'_-$ ($\Re\,z'<0$ half-plane) for the fractional propagator $(k^4)^{-\g/2}$. Here the two branches of the hyperbola are drawn with $\om$ fixed and two choices $|k^0|\gtrless\om$.}
\end{figure}

\emph{Step 2.} According to \Eqqs{Gbr0} and \Eq{Gbr1} and the discussion in \ref{appA}, the functions \Eq{Gprime} are
\be\label{Gprimeg}
\tilde G_\pm'(z',\om)= \frac{2z'}{\pm z'+k^0}\,\left[\mp(z^{\prime2}-\om^2)\right]^{-\g}\,.
\ee
They have a zero at $z'=0$, a pole at $z'=\mp k^0$, a branch-point singularity at $z'=\pm\om$ and a locus of discontinuity given by a hyperbola with vertices $\pm\om$ and foci in $z'=\pm\sqrt{2}\,\om$ (Fig.~\ref{fig9}). Indeed, applying the transformations \Eq{transf} to \Eqq{Gtot}, the discontinuity $z=\rmi t$ on the imaginary axis with $t\in(-\infty,+\infty)$ becomes a hyperbola with branches $z_\pm'=\pm\sqrt{\om^2+\rmi t}$, i.e.,
\be
\frac{(\Re\,z')^2}{\om^2}-\frac{(\Im\,z')^2}{\om^2}=1\,.
\ee
When $t\to\pm\infty$, $z_+'\simeq \sqrt{t}\,\rme^{\pm\frac{\rmi\pi}{4}}$ and $z_-'\simeq \sqrt{t}\,\rme^{\pm\frac{\rmi\pi}{4}+\rmi\pi}$, which are the asymptotes at $\pm\pi/4$ (mapping of $\G_+$ via $z_+'$) and $3\pi/4$, $5\pi/4$ (mapping of $\G_-$ via $z_-'$). The arcs at infinity in Fig.~\ref{fig8} are transformed into arcs at infinity of angle $\pi/2$ comprised between the two asymptotes; these arcs are thrown away eventually. Each mapping $z_\pm'$ leads to a different situation where the poles $\pm k^0$ lie inside or outside the hyperbola, depending on whether $|k^0|\gtrless\om$ (Fig.~\ref{fig9}). The mapping via $z_+'$ covers the right half of the plane and has $k^0>\om$, so that $z_+'=k^0$ lies inside the hyperbola. On the other hand, the mapping via $z_-'$ covers the left half of the plane and has $|k^0|<\om$, so that $z_-'=-k^0$ lies outside the hyperbola. This ``jump'' of the pole across the discontinuity locus is the result of the rigid mapping of $\G_\pm$ through the map $z_\pm(z)$, which are not homotopic deformations.

\emph{Step 3.} The Cauchy integrals in $z_\pm'$ identify $k^0=\pm\om$ as the singularities of the propagator, which are actually branch points from which the branch cuts $k^0>\om$ and $k^0<-\om$ originate. The hyperbola in the $z'$-plane is nothing but the integration domain of $k^0$, i.e., the path $\G^0$ we are looking for.

\emph{Step 4.} Finally, we deform homotopically the hyperbola so that to maximize contact with the real axis in the complex $k^0$-plane. Each hyperbolic branch is wrapped around one of the branch cuts and we obtain exactly the path $\G_+^u\cup\G_-^u$ of Fig.~\ref{fig3}.



\section{Solution of the diffusion equation \Eq{difG}}\label{appD}

In this appendix, we find the solution of the diffusion equation \Eq{difG} with Lorentzian kinetic operator:
\be\label{difGapp}
(\p_r+\lst^4\B^2_x)\cG(r\lst^4,x-x')=0\,,\qquad \cG(0,x-x')=\de^D(x-x')\,.
\ee
Since similar heat equations in the literature all have a Euclidean kinetic term, we cannot borrow past results and must calculate $\cG(r,x-x')$ anew.

In momentum space, the diffusion equation \Eq{difGapp} is
\be\label{difGk}
(\p_r+\lst^4k^4)\tilde\cG(r\lst^4,k)=0\,,\qquad \tilde\cG(0,k)=1\,,
\ee
where $k^4=(k^2)^2$. The solution is
\be\label{gk4}
\tilde\cG(r\lst^4,k)=\rme^{-r\lst^4k^4}\,.
\ee
Set $x'=0$ for simplicity. In $D=4$ dimensions, the Fourier anti-transform is
\be\label{fouan}
\cG(r\lst^4,x)=\int\frac{\rmd^4k}{(2\pi)^4}\,\rme^{\rmi k\cdot x}\tilde\cG(r\lst^4,k)\,.
\ee
Calling $k=|\bm{k}|$ and $R=|\bm{x}|$, we have $\rmd^4k=\rmd k^0\,\rmd k\,k^2\,\rmd\cos\vartheta\,\rmd\vp$, $\rmi k\cdot x=-\rmi k^0 t+\rmi kR\cos\vartheta$, $\vp\in[0,2\pi]$, $\cos\vartheta\in[-1,1]$, $k^0\in (-\infty,+\infty)$, so that
\ban
\cG(r\lst^4,x)&=&\frac{2}{(2\pi)^3}\int\rmd k^0\int_0^{+\infty}\rmd k\,\frac{k\,\sin(kR)}{R}\,\rme^{-\rmi k^0 t}\rme^{-r\lst^4 k^4}\\
&=&\frac{1}{(2\pi)^3}\frac{1}{\rmi R}\int\rmd k^0\int_{-\infty}^{+\infty}\rmd k\,k\,\rme^{-\rmi k^0 t+\rmi k R}\rme^{-r\lst^4 k^4}.
\ean
Introducing the light-cone variables $k^\pm\coloneqq k^0\pm k$ and $x^\pm\coloneqq t\pm R$, we get
\ba
\cG(r\lst^4,x)&=&\frac{1}{(2\pi)^3}\frac{1}{4\rmi R}\int\rmd k^+\,\rmd k^-\,(k^+-k^-)\,\rme^{-\frac{\rmi}{2} k^+ x^--\frac{\rmi}{2} k^- x^+}\rme^{-r\lst^4 (k^+)^2(k^-)^2}\nn
&=&\frac{1}{(2\pi)^3}\frac{1}{2R}\left(\p_--\p_+\right)\int\rmd k^+\,\rmd k^-\,\rme^{-\frac{\rmi}{2} k^+ x^--\frac{\rmi}{2} k^- x^+}\rme^{-r\lst^4 (k^+)^2(k^-)^2}\nn
&=& \frac{1}{(2\pi)^3}\frac{1}{2R}\left(\p_--\p_+\right)\frac{\pi}{\sqrt{r\lst^4}}\,G_{03}^{20}\left(z~\bigg|~\begin{matrix}\\0,0,\frac12\end{matrix}\right),\quad z=\frac{(x^+x^-)^2}{256\,r\lst^4}\,,
\ea
where $G_{03}^{20}$ is the Meijer G-function defined by the Mellin--Barnes representation \cite[Section 9.3]{GR}
\be
G_{03}^{20}\left(z~\bigg|~\begin{matrix}\\0,0,\frac12\end{matrix}\right)=\frac{1}{2\pi\rmi}\int_{\G_G}\rmd \varrho\,z^\varrho\,\frac{\G^2(-\varrho)}{\G\left(\frac12+\varrho\right)}\,.
\ee
The contour $\G_G$ encircles the positive semi-axis $\Re\,\varrho\geq 0$ clockwise \cite[chapter 16]{NIST} and, therefore, all the poles of $ \Gamma^2 (-\varrho)$. Due to the asymptotic behaviour of the gamma functions (Stirling formula), quarter circles at infinity in the $\Re\,\varrho\geq 0$ half-plane do not contribute to the Mellin--Barnes integral and the contour $\G_G$ can be opened and continuously deformed to the imaginary $\varrho$ axis, keeping the origin $\varrho=0$ to its right. Since
\ben
\left(\p_--\p_+\right)z=\frac{(x^+x^-)R}{64\,r\lst^4}\,,
\een
we have
\be
\cG(r\lst^4,x)= \frac{x^+ x^-}{2^{10}\pi^2 (r\lst^4)^{\frac32}}\frac{\rmd}{\rmd z}\,G_{03}^{20}\left(z~\bigg|~\begin{matrix}\\0,0,\frac12\end{matrix}\right)= -\frac{x^+ x^-}{2^{10}\pi^2 (r\lst^4)^{\frac32}}\,G_{03}^{20}\left(z~\bigg|~\begin{matrix}\\-1,0,-\frac12\end{matrix}\right),\label{meijG}
\ee
where $x^+ x^-=-s^2=t^2-R^2$ is the finite line element; restoring $x'\neq 0$, we reach \Eq{solG}. In this case, the Mellin--Barnes representation of the Meijer G-function is 
\be
G_{03}^{20}\left(z~\bigg|~\begin{matrix}\\-1,0,-\frac12\end{matrix}\right)=\frac{1}{2\pi\rmi}\int_{\G_G}\rmd \varrho\,z^\varrho\,\frac{\G(-\varrho) \G(-\varrho-1)}{\G\left(\frac32+\varrho\right)}\,,
\ee
and the contour $\G_G$ goes from $-\rmi \infty$ to $+\rmi\infty$ and indents the pole at $\varrho= -1$ to the right of $\G_G$ (alternatively, $\G_G$ can be continuously deformed in such a way as to encircle the positive semi-axis $\Re\,\varrho\geq -1$ clockwise). The Meijer G-function above is singular at $z=0$, as expected on general grounds. Its behaviour can be easily obtained by evaluating the first residues at the poles at $\varrho=-1,0,\dots$:
\be\label{smalz}
G_{03}^{20}\left(z~\bigg|~\begin{matrix}\\-1,0,-\frac12\end{matrix}\right)= \frac{1}{\sqrt{\pi}}\left(\frac{1}{z} + 2 \ln z + \cdots \right),
\ee
where the dots denote regular terms at $z=0$.

Equation \Eq{meijG} is is the generator of the most general solution of the diffusion equation, according to \Eq{solgen}. To make an analogy, the delta function in \Eq{difGapp} corresponds to the initial condition of a droplet of ink in water, while \Eq{meijG} is the distribution of the ink at time $r$. The analogy stops here because $r$ is an artificial parameter and $\cG$ is a spacetime distribution but there are still some similarities with respect to a standard diffusion process governed by Brownian motion. For example, the $(1+1)$-dimensional diffusion equation for Brownian motion is $(\p_t-\lst\p_x^2)\cG(t,x-x')=0$, where $\lst$ is the length diffusion coefficient; the Dirac distribution $\de(x-x')$ evolves into a Gaussian $\cG(t,x-x')=\exp[-(x-x')^2/(4\lst t)]/\sqrt{2\pi \lst t}$. The behaviour of \Eq{meijG} is not very different, since, at any finite $r$, the function $\cG\propto \sqrt{z}\, G_{03}^{20}\sim 1/\sqrt{z}$ blows up at $s=s'$ (on the light-cone) according to \Eq{smalz} and it decays at large $|s-s'|$. This spill-over of the support from the light-cone remains after integrating on $\tau$ to get the Green's function of the operator $\Bd^\g$ and is a well-known violation of Huygens' principle typical of fractional d'Alembertians \cite{Gia91,doA92,BGO,BG}.

Expression \Eq{meijG} is similar but essentially new with respect to solutions of other quartic diffusion equations known in the literature of transport theory and mathematics. These equations are Euclidean versions of \Eq{difGapp} with different choices of sign in front of the kinetic term and, possibly, with extra sources.

A widely studied case is the parabolic diffusion equation with quartic Euclidean kinetic operator \cite{Hoc78,HO,BO2}
\be\label{N4eq1}
(\p_r+\N^4)u(r,x)=0\,,\qquad u(0,x)=u_0(x)\,.
\ee
Expressed as a Fourier anti-transform, its solution $u(r,x)$ has the same form as our \Eq{gk4} and \Eq{fouan} but with Euclidean $k^2$.

Another model, somewhat more artificial in transport theory but used in image processing \cite{PeMa,YXTK,YoKa}, is the hyperbolic diffusion equation with quartic Euclidean kinetic operator \cite{Fun79}
\be\label{N4eq2}
(\p_r-\N^4)u(r,x)=0\,,\qquad u(0,x)=u_0(x)\,,
\ee
solved in one dimension as the limit of a yet different diffusion equation \cite{HO}.
A similar equation but with a non-trivial source term, 
\be\label{N4eq3}
(\p_r-\N^4)u(r,x)=\frac{\N^2u_0(x)}{\sqrt{\pi r}}\,,
\ee
arises in the context of iterated Brownian motion when applying twice the fractional heat equation $(\p_r^{1/2}-\N^2)u(r,x)=0$ \cite{OZ,AlZh,All02,DeB04,BMN1,BOS} (see \cite{MeK,Calcagni:2012rm} for reviews). Equation \Eq{N4eq2} is a special case of \Eq{N4eq3} for $u_0(x)=0$. The solution is a Meijer G-function similar to \Eq{meijG} but with different indices and $(s-s')^2$ replaced by the Euclidean radius.


\end{document}